\def\jobname{craterleo}
\documentclass{aa}  

\usepackage{amsmath}
\usepackage{gensymb}

\usepackage{graphicx}
\usepackage{txfonts}
\usepackage{float}
\usepackage{array}
\usepackage{physics}
\usepackage{amsfonts}
\usepackage{amssymb}
\usepackage{xcolor}
\usepackage{placeins}
\usepackage{multirow}
\usepackage{soul}
\usepackage{longtable}
\usepackage{rotating}

\begin{document}

   \title{Satellite group infall into the Milky Way: exploring the Crater-Leo case with new HST proper motions}


   \author{Mariana P. Júlio \inst{1, 2} \and Marcel S. Pawlowski \inst{1} \and Sangmo Tony Sohn \inst{3} \and Salvatore Taibi \inst{1} \and Roeland P. van der Marel \inst{3,4} \and Stacy S. McGaugh \inst{5}}

   \institute{Leibniz-Institut für Astrophysik Potsdam (AIP), An der Sternwarte 16, D-14482 Potsdam, Germany \and Institut für Physik und Astronomie, Universität Potsdam, Karl-Liebknecht-Straße 24/25, D-14476 Potsdam, Germany \and Space Telescope Science Institute, 3700 San Martin Drive, Baltimore, MD 21218, USA \and William H. Miller III Department of Physics \& Astronomy, Johns Hopkins University, Baltimore, MD 21218, USA \and Department of Astronomy, Case Western Reserve University, 10900 Euclid Avenue, Cleveland, OH 44106, USA}

   \date{Received 14 March 2024 / Accepted 23 April 2024}

 
  \abstract
   {Within $\Lambda$ Cold Dark Matter ($\Lambda$CDM) simulations, Milky Way-like galaxies accrete some of their satellite galaxies in groups of 3-5 members rather than individually. It was also suggested that this might be the reason behind the origin of satellite planes. Objects accreted in groups are expected to share similar specific total energy and angular momentum, and also identical orbital planes and directions.}
   {Looking at observations of Milky Way satellites, the dwarf galaxies Leo II, IV, V, Crater II, and the star cluster Crater 1 were proposed to be a vestige of group infall. The suggested "Crater-Leo group" shows a coherent distance gradient and all these objects align along a great circle on the sky. We use proper motion data to investigate whether the phase-space distribution of the members of the proposed group are indeed consistent with group infall.} 
   {To further investigate this possibility, we use \textit{Gaia} Data Release 3 (DR3) and new Hubble Space Telescope (HST) proper motions, $(\mu_{\alpha*}, \mu_\delta) = (-0.1921 \pm 0.0514, -0.0686 \pm 0.0523)$ mas yr$^{-1}$ for Leo IV and $(\mu_{\alpha*}, \mu_\delta) = (0.1186 \pm 0.1943, -0.1183 \pm 0.1704)$ mas yr$^{-1}$ for Leo V, to derive accurate orbital properties for the proposed group objects. In addition, we explore other possible members of this putative association. \par}
   {Leo II, Leo IV, and Crater 1 show orbital properties consistent with those we predict from assuming group infall. However, our results suggest that Crater II was not accreted with the rest of the objects. If confirmed with increasingly accurate proper motions in the future, the Crater-Leo objects appear to constitute the first identified case of a cosmologically expected, typical group infall event, as opposed to the highly hierarchical Magellanic Cloud system.}
   {}


   \maketitle
%

\section{Introduction}

$\Lambda$-cold dark matter ($\Lambda$CDM), the standard model of cosmology, predicts that structure grows hierarchically. This means that small objects are formed first and eventually collapse under the influence of their own gravity, merging with each other to form more massive and larger objects (e.g. \citealp{white1978}, \citealp{blumenthal}).
In this scenario, the small objects - satellites - can orbit their host halo, but they will ultimately merge with or be tidally disrupted by it (e.g. \citealp{gao_subhalo_2004}). Looking at the formation and evolution of galaxies in $\Lambda$CDM simulations, it was found that a large number of these satellites survived to the present day (e.g. \citealp{springel_aquarius_2008}) and that many of them were accreted in groups onto Milky Way-like halos, a scenario \citet{lynden-bell_dwarf_1976} proposed before the advent of $\Lambda$CDM.  

Indeed \citet{li_infall_2008} find in simulations\footnote{Since they use dark matter only simulations, their results are not strongly limited by resolution.} that at least 1/3 of the present-day satellites have fallen in groups and typically these groups consist of 3 to 5 satellites of similar masses, contrary to the Magellanic Cloud (MC) system. \citet{wheeler_sweating_2015} shows that low-luminosity galaxies are expected to host luminous galaxies of their own, by using high-resolution hydrodynamical simulations. \citet{Wetzel2015} estimates significant group infall fractions ($30\%-60\%$). This would mean that 18-36 of the known $\approx 60$ satellite galaxies of the Milky Way (MW) should have been part of some group before being accreted onto its halo. If we also consider the globular clusters and assume a similar fraction, this number increases to 60-120 of the total objects ($\approx 210$). \citet{Shao2018} uses two different types of hydrodynamical simulations to study group and filamentary dwarf galaxy accretion into Milky Way (MW) mass haloes. They find that only $14\%$ of the present-day 11 most massive satellites were accreted in pairs, $14\%$ in triplets and higher groups were extremely unlikely. However, these numbers increase for fainter satellites, with $12\%$ being accreted in pairs, and $28\%$ in richer groups. \citet{bakels_pre-processing_2021} uses high-resolution dark-matter-only simulations and finds that 20$\%$ of all the accreted systems were accreted in groups. 

This process is often suggested as an important mechanism to produce spatial and kinematic coherence between satellite galaxies, such as the so-called planes of satellites (see e.g. \citealp{li_infall_2008} and \citealp{donghia_small_2008}). \citet{kunkel1976royal} and \citet{lynden-bell_dwarf_1976} first discovered that the satellite galaxies align along a polar great circle around the MW. More recently, it was found that the anisotropic distribution of these satellites, now called the Vast Polar Structure (VPOS, \citealp{pawlowski_vpos_2012}), also shows coherent motion, being inconsistent with the predictions of $\Lambda$CDM (\citealp{pawlowski_vpos_2012} and \citealp{pawlowski_milky_2020}). Similar anisotropic structures have also been discovered around Andromeda (M31), the Great Plane of Andromeda (GPoA, \citealp{koch_anisotropic_2006}, \citealp{mcconnachie_satellite_2006} and \citealp{ibata_vast_2013}) and around Centaurus A (\citealp{tully_two_2015}, \citealp{muller_testing_2016} and \citealp{muller_whirling_2018}). 

\citet{lynden_1982} first proposed associations between satellite galaxies and globular clusters in the Local Group. In their seminal paper, \citet{lynden-bell_ghostly_1995} show that satellites accreted together leave signatures of group infall in their phase-space distribution. It is expected that objects that were accreted together share similar specific angular momentum and total energy. Using this argument and assuming a spherical potential for the MW, they are able to identify several possible associations between the known satellites at the time and predict their proper motions. 

\citet{li_infall_2008} finds that infalling groups can remain coherent and share orbital planes for up to 8 Gyr, the typical time that satellites have been within MW/M31 halo, making this argument more compelling. The simulations performed by \citet{Wetzel2015} back up this finding. Group infall could also explain the high incidence of satellite galaxy pairs in the Local Group (see e.g. \citealp{fattahi_galaxy_2013}). 

In the $\Lambda$CDM paradigm, it is expected that halos of $\sim10^{11}$M$_\odot$ contain subhalos that have sufficient gravitational potential to host their own subhalos (satellites of satellite galaxies, \citealp{Patel2020}). In the Milky Way (MW), the Large Magellanic Cloud (LMC) is the only galaxy in the $\sim10^{11}$M$_\odot$ range. \citet{donghia_small_2008} and \citet{sales_clues_2011} suggested that the Magellanic Clouds (MCs) were the largest members of a group of dwarf galaxies that fell onto the MW halo, originating the majority of the brighter satellites of the MW. Later studies found that LMC should host $\approx 5-10$ ultra-faint dwarf galaxies with $M_\star \approx 10^2-10^5$M$_\odot$ (e.g. \citealp{dooley_predicted_2017} and \citealp{jahn_dark_2019}). Recently, proper motions of the lowest-mass satellite galaxies were measured for the first time (\citealp{Fritz2018_gaiadr2} and \citealp{pace_proper_2019}), allowing for the determination of which galaxies are dynamically associated with the MC \citep{kallivayalil_missing_2018}. \citet{Patel2020} identify 3-6 galaxies as MCs satellites, being consistent with the low end of the cosmological expectations. There is also observational evidence of low mass groups existing in isolated environments \citep{stierwalt_direct_2017}.

However, \citet{li_infall_2008} and \citet{Wang2013} find that the typical group infall events in simulations consist of objects of similar mass. We, therefore, expect other group associations among the satellites of the MW, but observational evidence for past associations has been, until now, absent. Several other associations were proposed based on their positions and velocities, such as Fornax, Leo I, Leo II and Sculptor \citep{lynden_1982}, later revisited with additional proposed members, Sextans, and Phoenix \citep{majewski_fornax-leo-sculptor_1994}; Pisces II and Pegasus III (\citealp{kim_heros_2015}, \citealp{garofalo_born_2021}, and \citealp{richstein_structural_2022}); or NGC 147, NGC 185 and Cassiopeia II in the M31 galaxy \citep{arias_ngc_2016}. \citet{tully_squelched_2002, tully_associations_2006} proposed several bound associations of dwarfs in the Local Group and beyond, with strong spatial and kinematic correlations, although they are of larger scales than the ones suggested between satellite galaxies. \citet{fattahi_galaxy_2013} found that $\sim 30\%$ of Local Group satellites, brighter than $M_\mathrm{V}=-8$, are likely in pairs, such as And I/And III and NGC 147/NGC 185 in the M31 galaxy, even though cosmological simulations predict that less than $4\%$ of satellites are part of pairs in this range of luminosities. The suggestion that NGC 147 and NGC 185 were likely associated was first proposed by \citet{van_den_bergh_binary_1998} due to their proximity in positions, distance and velocities, but \citet{sohn_hst_2020}, using HST proper motions, finds that it is very unlikely that these galaxies were ever gravitationally bound. \citet{bell_ultrafaint_2022} argues that the extremely asymmetric satellite distribution of M81 indicates that many of the satellites were recently accreted as a group. However, none of these suggestions, with the exception of the MCs system, was confirmed through proper motion measurements. Furthermore, group infall on MW-like systems has been studied almost exclusively in cosmological simulations but not observationally. \par

\citet{jong_enigmatic_2010} first proposed the association between Leo IV~\citep{belokurov_cats_2007} and Leo V~\citep{belokurov_leo_2008} to explain their proximity in distance and in radial velocity. Although they find that these galaxies do not share the same orbit when a spherical potential is assumed, they conclude that they most likely fell into the MW halo together. When Crater 1 was discovered \citep{belokurov_atlas_2014}, it was proposed as an additional possible member of the association. This possible association was further discussed by \citet{Torrealba2016}, with the discovery of Crater II. They additionally included Leo II \citep{harrington_two_1950} since all five objects lie close to the great circle with the pole at $(\alpha, \delta) = (83.2\degree, -11.8\degree)$. \par

With measured proper motions we can properly explore the suggested association between the objects in the Leo-Crater group. So to further investigate the proposed group, we present and derive here accurate proper motions of Leo IV and V using new data from the Hubble Space Telescope (HST), and combine these with existing \textit{Gaia} and HST measurements for the other objects in the group.
If these satellites are conclusively identified to have been accreted together, not only will they be the first confirmed case of group accretion besides the MCs system, but they can start to provide constraints on the environmental influences on dwarf galaxy formation and evolution in general, and the assembly history of the MW satellite system in particular.

In Sect.~\ref{sec:group} we describe the data and observations of the objects proposed as a group. This is followed in Sect.~\ref{sec:methods} by the methods that we used to understand if these objects were accreted together and by our approach to find other possible associations in Sect.~\ref{sec:pudim}. We continue in Sect.~\ref{sec:results} with our results on the predicted proper motions of the discussed objects (Sect.~\ref{sec:predicted-pm}), and their corresponding orbital properties (Sect.~\ref{sec:orbital-parameters}). We check for additional members of the group not considered before in Sect.~\ref{sub:additional-members} and if the infall of the LMC had any effect on the measured direction of the angular momentum of the objects in Sect.~\ref{sec:effect-lmc}. We end with a discussion and our conclusions in Sect.~\ref{sec:discussion}. In Appendixes~\ref{app:mcmc},~\ref{app:mwmass},~\ref{app:additional-members},~\ref{app:orbital} and~\ref{app:lmceffect}, we show supplementary material to support the results that we discuss in the main text. Throughout the paper, the data from \textit{Gaia} DR3 is always represented in shades of cold colours, and the data from HST is always represented in shades of warm colours to facilitate recognition and association.

\section{The proposed Crater-Leo group}\label{sec:group}
The proposed Crater-Leo group, represented in Figure~\ref{fig:positions}, consists of Leo II, a classical dwarf spheroidal galaxy, Leo IV and Leo V, both ultra-faint satellite galaxies, Crater 1, a globular cluster, and Crater II, a distant diffuse satellite galaxy.

\subsection{Properties}\label{subsec:properties}
\begin{figure*}
    \centering
    \includegraphics[scale=0.55]{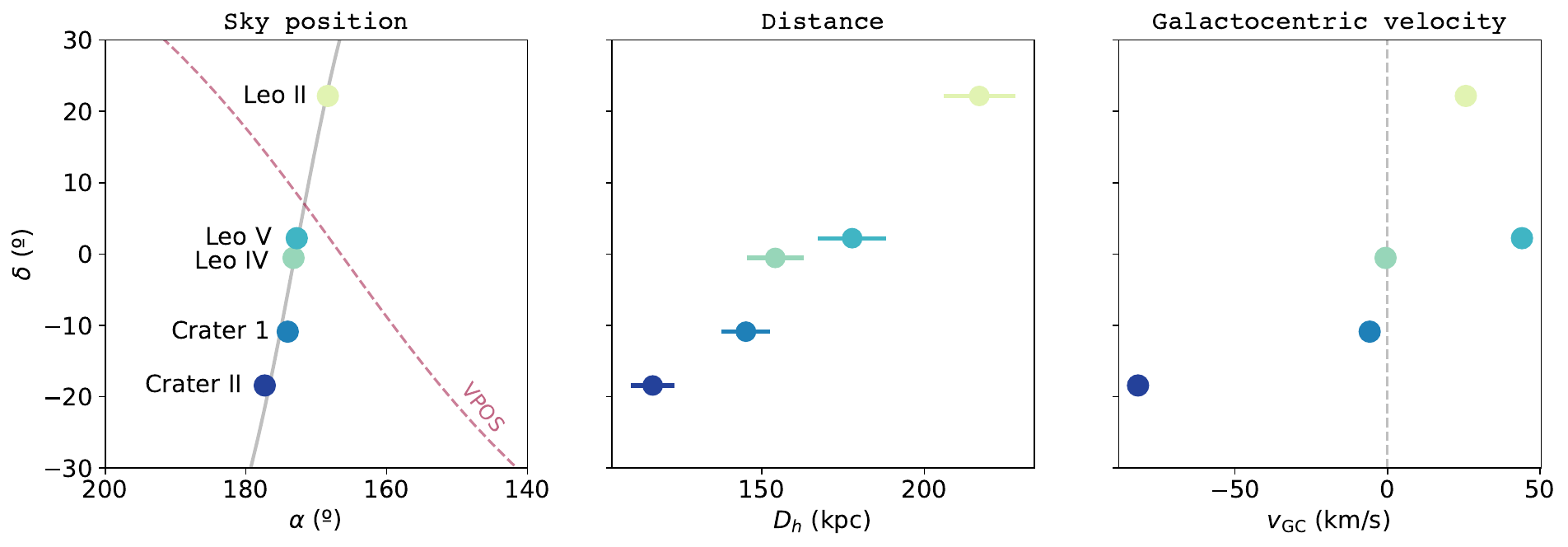}
    \caption{The Crater-Leo objects. The left-hand panel shows the positions of Leo II, Leo IV, Leo V, Crater 1, and Crater II in the sky. The solid grey line represents the great circle with the pole at $(\alpha, \delta) = (263.01\degree, 10.78\degree)$ that passes close to all five satellites. The red dashed line represents the best-fit orbital plane of the VPOS at $(\alpha, \delta) = (76.53\degree, 36.24\degree)$ estimated by \citet{Fritz2018_gaiadr2}. The middle panel represents the declination versus the heliocentric distance of the satellites. The right-hand panel shows the declination versus line-of-sight velocity of the satellites, with $v_{\mathrm{GC}} = 0$ kms$^{-1}$ represented by the dashed line.}
    \label{fig:positions}
\end{figure*}

Leo IV and Leo V were first proposed as a pair to explain both their spatial proximity ($<3\degree$) and small differences in their line-of-sight distances ($<20$ kpc) and velocities ($<40$ kms$^{-1}$) \citep{jong_enigmatic_2010}. Their closeness is especially unusual since the outer halo of the MW is mostly empty, with only $\sim 20\%$ of all currently known dwarfs inside a 400 kpc radius located beyond 150 kpc from the Galactic centre \citep{Torrealba2016}. Following the discoveries of Crater 1 \citep{belokurov_atlas_2014} and Crater II \citep{Torrealba2016}, the possible association between these objects was further explored. Together with Leo II, these objects align along a common great circle (see left panel of Figure~\ref{fig:positions}), which is highly unlikely. \citet{Torrealba2016} estimated the probability of this happening randomly - without sharing a common group origin - using random Monte Carlo realisations of satellite distributions on the sky and determining the largest group of satellites that are strongly aligned for each realisation. They find that having apparent aligned groups with five or more members by random chance has a probability of only $\sim 0.0045$.   \par 
In addition, the objects show a monotonic distance gradient (see middle panel of Figure~\ref{fig:positions}), moving up in declination from Crater II ($D_h = 116.6 \pm 6.6$ kpc) to Leo II ($D_h = 216.8 \pm 11.0$ kpc). To make this argument even more compelling, the line-of-sight velocities of these objects seem to prefer a common value of about zero in the Galactic rest frame\footnote{The root mean square (RMS) scatter is $\sim40$ km/s, which can be compared to the larger one-dimensional velocity dispersion of $\sim 100$ km/s for the MW satellite population as a whole \citep{riley_velocity_2019}. When we exclude Crater II, this RMS scatter becomes $\sim 25$ km/s.} (see right panel of Figure~\ref{fig:positions}), with Crater II being the most offset in velocity. \par 
 \par 

Furthermore, \citet{fattahi_galaxy_2013} show that satellite galaxies close in position and velocity preferentially form pairs of comparable luminosity. When we look at the luminosity of these objects, Crater 1, Leo IV, and Leo V have comparable luminosities ($M_\mathrm{V} = -5.3$, $M_\mathrm{V} = -5.8$ and $M_\mathrm{V} = -5.2$, respectively), as well as Leo II and Crater II ($M_\mathrm{V} = -9.8$ and $M_\mathrm{V} = -8.2$, respectively). These values are described in Table~\ref{tab:properties}.

\begin{table*}[ht]
	\centering
        \caption{Properties of the Crater-Leo objects: coordinates of the optical centre (RA and DEC) in degrees, absolute magnitude in the V-band ($M_\mathrm{V}$), the distance modulus (dm), heliocentric systemic line-of-sight velocity ($\langle v_\mathrm{los}\rangle$) in km s$^{-1}$, mean stellar metallicity ($\langle$[Fe/H]$\rangle$), dynamical/stellar mass ($M$) in solar masses, and the corresponding references.}

	\label{tab:properties}
	\begin{tabular}{lcrrcrrcl} 
		\hline
		Object     & RA (º)      & DEC (º)     & $M_\mathrm{V}$  & dm              & $\langle v_\mathrm{los}\rangle$ (km s$^{-1}$) & $\langle$[Fe/H]$\rangle$                  & $M^*$ $(10^6 M_\odot)$ & References         \\ 
            \hline
		Leo II     & $168.3627$ & $22.1529$  & $-9.8\pm 0.3$   & $21.68\pm 0.11$ & $78.30^{+0.60}_{-0.60}$  & $-1.63^{+0.01}_{-0.01}$ & 4.6 & 1, 2, 3, 4, 5, 6      \\
		Leo IV     & $173.2405$ & $-0.5453$  & $-5.8\pm 0.4$   & $20.94\pm 0.07$ & $131.40^{+1.20}_{-1.10}$ & $-2.47^{+0.14}_{-0.14}$ & 1.3 & 1, 7, 8, 9, 6      \\
		Leo V      & $172.7857$ & $2.2194$   & $-5.2\pm 0.4$   & $21.25\pm 0.08$ & $173.00^{+1.00}_{-0.80}$ & $-2.28^{+0.15}_{-0.16}$ & 1.1 & 1, 7, 10, 9, 6      \\
            Crater 1   & $174.0660$ & $-10.8778$ & $-5.3\pm 0.1$   & $20.81\pm 0.05$ & $149.30^{+1.20}_{-1.20}$ & $-1.68^{+0.05}_{-0.05}$ & 0.01 & 11, 12, 13 \\
            Crater II  & $177.3280$ & $-18.4180$ & $-8.2\pm 0.1$   & $20.33\pm 0.07$ & $87.60^{+0.40}_{-0.40}$  & $-1.95^{+0.06}_{-0.05}$ & 4.4 & 14, 15, 16, 17 \\
		\hline
	\end{tabular}
\par \vspace{4mm}
\small \raggedright {References: (1) \citet{munoz_megacam_2018}; (2) \citet{irwin_structural_1995}; (3) \citet{gullieuszik_evolved_2008}; (4) \citet{spencer_binary_2017}; (5) \citet{kirby_universal_2013}; (6) \citet{McConnachie2012}; (7) \citet{jong_enigmatic_2010}; (8) \citet{moretti_leo_2009}; (9) \citet{jenkins_very_2021}; (10) \citet{mutlu-pakdil_signatures_2019}; (11) \citet{belokurov_atlas_2014}; (12) \citet{weisz_hubble_2016}; (13) \citet{kirby_spectroscopic_2015}; (14) \citet{vivas_decam_2020}; (15) \citet{Torrealba2016}; (16) \citet{caldwell_crater_2017}; (17) \citet{fu_dynamical_2019}.} \par
\small {$^*$For the dwarf galaxies (Leo II, Leo IV, Leo V and Crater II), $M$ corresponds to the dynamical mass inside the half-light radius. For the globular cluster (Crater 1), due to its lack of dark matter, $M$ corresponds to its stellar mass.}

\end{table*}

Finally, some of these objects also seem to share similar star formation histories (SFHs). The SFHs of Leo II and Leo IV indicate that these objects ended their star formation around 5-6 Gyr ago \citep{weisz_star_2014}. Crater 1's stellar content appears to be metal-poor and old, with an age between 7 and 10 Gyr, with three stars younger than 1 Gyr \citep{belokurov_atlas_2014}. It has been argued before that the globular clusters (GCs) on the outskirts of the MW were accreted together with their parent dwarf galaxies (see e.g. \citealp{mackey_comparing_2004} and \citealp{huang_search_2021}). \cite{van_den_bergh_globular_2006} shows that the specific frequency of globular clusters fainter than $M_\text{V} \sim -7.5$ is particularly high in dwarf galaxies and that dwarfs with luminosity distributions similar to Leo II (with $M_V = -9.8\pm0.3$), can host globular clusters with luminosity distributions similar to Crater 1 (with $M_V = -5.3\pm0.1$)\footnote{For example, BKN 3N, with $M_V = -9.53$ has a globular cluster with $M_V = -5.23$ \citep{van_den_bergh_globular_2006}.}. \citet{weisz_hubble_2016} argues that Crater 1 joined the MW halo less than 8 Gyr ago. This makes plausible the idea that Crater 1 originated from Leo II since these objects share similar metallicities (see Table~\ref{tab:properties}). Leo II formed the second half of its stellar mass from 6 to 9 Gyr, which matches Crater 1's probable age.
The SFH of Leo V is not well determined; however, \citet{mutlu-pakdil_signatures_2019} argues that it should be similar to the SFH of Leo IV, since they have very similar colour-magnitude diagrams. 
The common end of star formation of the objects is consistent with an infall into the MW halo within the last 6 Gyr, with Crater 1 originating from one of them. Crater II, however, only shows signs of star formation events at 10.5 Gyr and 12.5 Gyr ago, with no presence of intermediate-age or younger stars \citep{walker_decam_2019}. It was found that this dwarf galaxy shows strong signals of tidal disruption, affecting its kinematics (see e.g. \citealp{fu_dynamical_2019} and \citealp{ji_kinematics_2021}). However, this does not affect the infered age of its stellar content. \par

\cite{li_infall_2008} found in cosmological simulations that accreted groups typically have 3-5 satellites, being consistent with the proposed group. However, when this group was first proposed, no proper motion measurements were available. With the measurements provided by \textit{Gaia} DR3, we can investigate the tentative group once again to confirm or disregard their association. \par 

\citet{Fritz2018_gaiadr2} explored this possibility by determining the possible orbital poles for the five objects with \textit{Gaia} DR2 proper motions. However, they found the association unlikely, since the orbital poles do not overlap well and had large uncertainties. They acknowledged the possibility of having two different groups (Crater 1 and Crater II as one of them, and Leo IV and Leo V as the other) instead of only one. However, they found that the orbital properties of Crater 1 and Leo II are consistent with each other, and Leo II also has orbital poles consistent with it being part of the other group, so they require better proper motions to confirm or exclude this possibility. \par 

We use the metric defined by \citet{geha_local_2010} to test whether the Crater-Leo objects are currently bound to each other as was done in \citet{geha_hstacs_2015} and \citet{sohn_hst_2020} for NGC 147 and NGC 185, and in \citet{richstein_structural_2022} for Pisces II and Pegasus III. Specifically, for two point masses to be gravitationally bound, their gravitational potential energy has to exceed their kinetic energy \citep{davis_abell_1995}, leading to the criterion $b \equiv 2GM_\mathrm{sys}/\Delta r\Delta v^2$, where $\Delta r$ is the total physical separation between the objects and $\Delta v$ the radial velocity difference between them. When $b > 1$, the system is considered bound. If our objects are indeed bound, we can assume they are associated. However, if they are not, this is not sufficient to disregard the possibility of them having been accreted together, since they might have just dispersed along their orbits. We determine this metric for every pair in our group, using the values described in Table~\ref{tab:properties} and the masses described in Table~\ref{tab:properties}, and we always get $b < 1$, meaning that none of the satellites are bound at the present time. For our satellites to be bound, taking into account their physical separation and the radial velocity difference, the mass of each pair would have to be $>10^{9} M_\odot$. We then need to perform an extensive analysis with up-to-date data, in particular recent and new proper motion measurements, to try to understand if these objects were accreted as a group.

\subsection{Proper motions}\label{subsec:propermotions}

With up-to-date data, described in Tables~\ref{tab:properties} and~\ref{tab:proper-motions}, we can expand upon this preliminary analysis again to test if the Crater-Leo objects are consistent with their expected dynamics if they were once part of a common group of satellites. We estimate the expected angular momentum of the group and its direction. With this, we can predict their proper motions and compare them with the current observations. This, combined with their orbital history, allows us to better assess the possibility of these objects being accreted together. \par 

The large uncertainties of the proper motions of Leo IV and Leo V, however, make this analysis more difficult. For that reason, we started and executed an observational project to determine improved proper motions from HST, which we present and discuss next.

\begin{table}[ht]
	\centering
	\caption{Proper motions ($\mu_{\alpha*}$ and $\mu_{\delta}$) in mas yr$^{-1}$ using \textit{Gaia} DR3 from \citet{Battaglia2022_gaiadr3}.}
	\label{tab:proper-motions}
	\begin{tabular}{lrr} 
		\hline
		Object & $\mu_{\alpha*}$ (mas yr$^{-1}$) & $\mu_{\delta}$ (mas yr$^{-1}$) \\
		\hline
		Leo II & $-0.11^{+0.03}_{-0.03}$ & $-0.14^{+0.03}_{-0.03}$   \\
		Leo IV & $-0.03^{+0.14}_{-0.14}$ & $-0.28^{+0.11}_{-0.12}$   \\
		Leo V  & $0.10^{+0.21}_{-0.21}$  & $-0.41^{+0.15}_{-0.15}$   \\
            Crater 1 & $-0.04^{+0.12}_{-0.12}$ & $-0.12^{+0.10}_{-0.10}$  \\
            Crater II & $-0.07^{+0.02}_{-0.02}$ & $-0.11^{+0.01}_{-0.01}$ \\
		\hline
	\end{tabular}
\end{table}

\subsubsection{New HST proper motions of Leo IV and Leo V}

Imaging data used for measuring proper motions of Leo~IV and Leo V were obtained with HST in two separate epochs. For Leo~IV, the first-epoch data were obtained on January 2012 through HST program GO-12549 (PI: T. Brown) in F606W and F814W for studying the star formation histories of Ultra-Faint Dwarf galaxies \citep{Brown2014}, and the second-epoch data were obtained on November 2016 through HST program GO-14236 (PI: S. T. Sohn) in F606W for measuring proper motions of Leo~IV along with other UFDs in the original \cite{Brown2014} sample. For Leo~V, the first- and second-epoch data were obtained on March 2017 (in F606W and F814W) and March 2020 (in F606W), respectively, through the multi-cycle HST program GO-14770 (PI: S. T. Sohn). 

To measure the PMs of both galaxies, we followed the methods used in our previous work on M31, and its two dwarf satellite galaxies NGC~147 and NGC~185 \citep{Sohn+2012, sohn_hst_2020}. We downloaded the {\tt \_flc.fits} images from the Mikulski Archive for Space Telescopes (MAST). For each star in each galaxy, we measured its position and flux from the {\tt \_flc.fits} images using the {\tt hst1pass} code \citep{Anderson+2022}. We corrected the positions for geometric distortion using the solutions by \citet{Kozhurina-Platais+2015}. For each dwarf galaxy, we then created high-resolution stacked images using images from the first-epoch data.

As the next step, we constructed color-magnitude diagrams (CMDs) for both Leo~IV and Leo~V using multi-band images obtained during the first epoch. The CMDs were used for identifying stars associated with each dwarf galaxy. We also identified background galaxies to be used as astrometric reference sources from the stacked images, first through an objective selection based on the quality-of-fit parameter output from the {\tt hst1pass} code, and then by visually inspecting each source.  We then constructed a template for each star or background galaxy by supersampling the scene extracted from the high-resolution stack. Templates constructed this way take into account the point-spread functions (PSFs), the galaxy morphologies, and the pixel binning. We fitted these templates to stars and galaxies in each individual {\tt \_flc.fits} image for measuring their positions. For the first epoch, we fitted the templates directly onto the stars and galaxies in the individual images. For the second epoch, we derived convolution kernels by comparing PSFs of numerous bright and isolated stars between the first and second epochs, and applied these kernels when fitting templates to allow for differences in PSF between different epochs. Once the template-fitting process was complete, we were left with the positions of all stars and background galaxies in each individual image in each epoch. 

The reference frames were defined by averaging the positions of stars from repeated first-epoch exposures. We used the positions of stars in each exposure of the second epoch to transform the positions of the galaxies into the reference frames. We then measured the positional difference for each background galaxy between the first and second epochs relative to the stars associated with Leo~IV and Leo~V. During this process, we applied ``local corrections’’ by making measurements only relative to stars with similar brightness in the local vicinity on the image to correct for residual CTE and any remaining geometric distortion systematics. We took the error-weighted average of all displacements of background galaxies with respect to Leo~IV and Leo~V stars for each individual second-epoch exposure to obtain an independent PM estimate.  We then obtained the final PMs of Leo~IV and Leo~V by computing the error-weighted mean of these individual PM estimates, multiplying $-1$ to this value to account for the fact that what we have measured is the reflex motion of background galaxies, and dividing by the time baseline in years. The final PMs and associated uncertainties derived as described in this subsection are listed in Table~\ref{tab:hst-proper-motions}.

\begin{table}[ht]
	\centering
	\caption{Proper motions ($\mu_{\alpha*}$ and $\mu_{\delta}$) in mas yr$^{-1}$ of Leo~IV and Leo~V from new HST measurements.}
	\label{tab:hst-proper-motions}
	\begin{tabular}{lrr} 
		\hline
		Object & $\mu_{\alpha*}$ (mas yr$^{-1}$) & $\mu_{\delta}$ (mas yr$^{-1}$) \\
		\hline
		Leo IV & $-0.1921 \pm 0.0514$ & $-0.0686 \pm 0.0523$   \\
		Leo V  & $0.1186 \pm 0.1943$  & $-0.1183 \pm 0.1704$   \\
		\hline
	\end{tabular}
\end{table}

For completeness, we also use the available HST proper motion for Leo II. \cite{piatek_proper_2016} measured this proper motion based on imaging with HST and Wide Field Camera 3 and got $(\mu_\alpha, \mu_\delta) = (-0.069\pm0.037, -0.087\pm0.039)$ mas yr$^{-1}$. With this result, they confirmed that the Leo II proper motion was aligned with the great circle that passes through Leo IV, Leo V, Crater 1 and Crater II, as first suggested by \cite{Torrealba2016}. They also estimate the pericentre of Leo II, since if they are orbiting in co-planar orbits with a range of energies (possibly acquired during the disruption of the group ), this value should be at least as small as the smallest current Galactocentric distance of the members of the group (120 kpc for Crater II). With their measured proper motion, they find that Leo II has a $31\%$ chance of having its pericentre at 120 kpc or less. Their results then support the hypothesis that Leo II, Leo IV, Leo V, Crater 1 and Crater II fell into the Milky Way as a group, with the direction of motion indicating that Leo II would be leading this group. \par 

\begin{figure}[ht]
    \centering
    \includegraphics[width=\columnwidth]{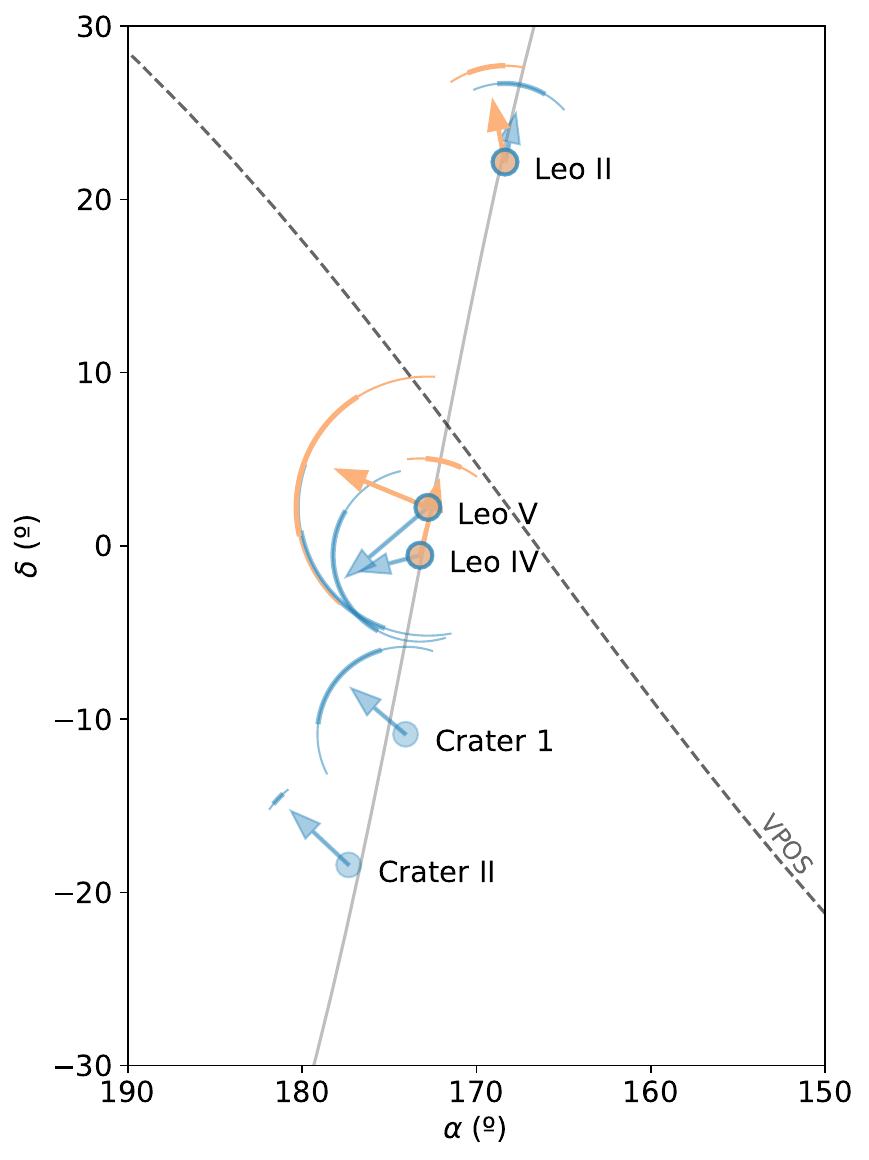}
    \caption{The positions of Leo II, Leo IV, Leo V, Crater 1, and Crater II in the sky. The solid grey line represents the great circle with the pole at $(\alpha, \delta) = (263.01\degree, 10.78\degree)$ that passes close to all five satellites. The black dashed line represents the best-fit orbital plane of the VPOS at $(\alpha, \delta) = (76.53\degree, 36.24\degree)$ estimated by \citet{Fritz2018_gaiadr2}. The arrows indicate the most likely direction of motion based on current proper motion estimates (the fainter ones, in blue, represent the \textit{Gaia} measurements, and the solid ones, in orange, the HST proper motions) and the circle segments indicate the 1 and 2$\sigma$ uncertainties in the velocity directions of the satellites.}
    \label{fig:group-pm}
\end{figure}

Figure~\ref{fig:group-pm} shows \textit{Gaia} DR3 proper motions represented by the blue arrows and the HST proper motions represented by the orange arrows, after being transformed into the Galactocentric frame. The arcs indicate $1 - 2\sigma$ intervals in the direction of motion. One can see that when we consider our HST proper motions for Leo IV and Leo V, their directions better align with the rest. However, since we do not know which one is closer to the true value, we will use both measurements in this study and discuss their implications for our results. Crater II's proper motion is well-constrained, thanks to the abundance of member stars measured by Gaia, and does not seem to align with the great circle that passes close to all five satellites. The other objects, however, seem to be consistent with this great circle within $2\sigma$, with almost all of them pointing in the same direction.

\section{Methods}\label{sec:methods}
We start by describing how we estimated the orbital poles of the Crater-Leo objects. This is followed by the explanation of the method that we use to determine if these objects were accreted together in Sect.~\ref{sec:pudim}. Finally, in Sect.~\ref{sec:pudim-pt2}, we describe how we determine possible additional members to the initially proposed group. 

As previously mentioned, if satellites are accreted together, we expect them to share similar specific angular momentum. Hence, they need to share orbital poles, i.e., the directions of the angular momentum vectors. To determine the uncertainties in the derived orbital poles, we used Monte Carlo realisations, and incorporated the uncertainties of the proper-motion measurements, of the distance, position, and line-of-sight velocity of the satellites (described in Table~\ref{tab:properties}), of the distance of the Sun from the Galactic centre, $D_\odot = 8.178\pm0.022$ \citep{gravity_collaboration_geometric_2019}, of the circular velocity of the local standard of rest (LSR), $v_\mathrm{circ} = 234.7\pm1.7$ \citep{nitschai_dynamical_2021}, and of the peculiar motion with respect to the LSR, $\vec{v}_\mathrm{LSR} = (11.10\pm0.72, 12.24\pm0.47, 7.25\pm0.37)$ \citep{schonrich_local_2010}.

\subsection{Lynden-Bell method}\label{sec:pudim}
\citet{lynden-bell_ghostly_1995} show that satellites that were accreted as a group share similar specific angular momentum $h$ and total energy $E$ and they use this argument to identify possible associations between the satellites of the MW at the time. By assuming a spherical and static potential of the form $\psi = -V^2_0 \ln r$, they are able to predict the proper motions of the objects they consider associated. A detailed explanation of its implementation can be found in their paper which we briefly summarise here for the reader's convenience. \par 

The total specific energy of a satellite's motion around the Milky Way at Galactocentric position $r=(x, y, z)$ and velocity $v=(v_x, v_y, v_z)$ can be written as
\begin{equation}\label{eq:totalenergy}
    E(r,v) = \frac{1}{2}v^2+\Phi_\mathrm{MW}(r),
\end{equation}
where $\Phi_\mathrm{MW}$ is the Galactic potential at the position of the satellite, and $v^2$ the total velocity of a satellite, given by $v^2 = v^2_\mathrm{r} + v^2_\mathrm{tan}$. 
Since the objects under study are at large distances, the vector pointing from the Galactic Centre to the position of the object is similar to their line-of-sight, so we can approximate the radial velocity that would be seen from the Galactic Centre $v_\mathrm{r}$ to the line-of-sight velocity $v_\mathrm{los}$. The tangential velocity $v_\mathrm{tan}$ relative to the Galactic Centre is unknown; however, we can introduce the specific angular momentum $h = |r\times v| = |r|\cdot |v_\mathrm{tan}|$ and rewrite Equation~\ref{eq:totalenergy} as
\begin{equation}
    E \approx \frac{1}{2}v_\mathrm{los}^2 + \frac{1}{2}h^2r^{-2} + \Phi_\mathrm{MW}(r).
\end{equation}
We can then write the radial Energy, $E_r$, as
\begin{equation}\label{eq:Er}
    E_\mathrm{r} = \frac{1}{2}v_\mathrm{r}^2+\Phi_\mathrm{MW}(r) = E-\frac{1}{2}h^2r^{-2}.
\end{equation}
Since objects that are associated should share the same $E$ and $h$ once both quantities are conserved, when we plot $E_r$ against $r^{-2}$, they will lie on a line of gradient $-g=-h^2/2$ and intercept $E$ \citep{lynden-bell_ghostly_1995}.

If we consider $\vec{s}$ the vector along the line-of-sight, $\vec{r}$ the Galactocentric position and $\vec{r_\odot}$ the position of the Sun, we have
\begin{equation}
    \vec{s} = \vec{r}-\vec{r_\odot}.
\end{equation}
The velocity of the object in the Galactocentric system of rest is then 
\begin{equation}\label{eq:vtotal}
    \vec{v} = v_r\hat{r}+\vec{h}\times\hat{r}/r.
\end{equation}
The radial velocity becomes
\begin{equation}\label{eq:vr}
    v_r = (v_l-\vec{h}\cdot\hat{r}\times\hat{s}/r) = (v_l-hr^{-1}\Delta)/\hat{r}\cdot\hat{s},
\end{equation}
where $v_l$ is the line-of-sight velocity corrected by the velocity of the Sun, $\vec{h}=h\hat{p}$ and $\Delta = \hat{p}\cdot(\hat{r}\times\hat{s})$.
To determine the magnitude and sign of $h$, we start by using an approximated value $h_0$ estimated by fitting Equation~\ref{eq:Er} to our data. The true value of the specific angular momentum is then given by $h_\pm = \delta h\pm h_0$. Using Equation~\ref{eq:vr} in Equation~\ref{eq:Er}, we get
\begin{equation}\label{eq:newEr}
    E_r = \frac{1}{2}[v_l-(\delta h\pm h_ 0)r^{-1}\Delta ]^2/(\hat{r}\cdot\hat{s})^2 - \Phi_\mathrm{MW} = E-\frac{1}{2}h^2r^{-2}.
\end{equation}
To find a better value for the specific angular momentum, we iteratively redetermine $-\frac{1}{2}h^2$ from the best-fitting line. This converges rapidly to give consistent values of $\pm h$. After five iterations, this value changes by less than $\approx 10^{-5}\%$, but we use 10 iterations just to be safe. For each possible association, we have two possible signs, however, the sign must be the same for all the members of a group, since it determines the direction of the orbit's circulation. 

Using Equations~\ref{eq:vtotal} and~\ref{eq:vr}, the total velocity relative to the Galactic centre is given by
\begin{equation}\label{eq:vfinal}
    \vec{v} = (v_l-hr^{-1}\Delta)\hat{r}/\hat{r}\cdot\hat{s}+\vec{h}\times\hat{r}/r.
\end{equation}
Having the expected 3D velocity determined, we are able to calculate the corresponding proper motions of the associated objects for each direction of $h$.

Here, we use their method to predict the proper motions that the Crater-Leo objects would have if they were associated. However, instead of using the spherical potential that they adopted, we use the gravitational potential described by \citet{bovy_galpy_2015} (\texttt{MWPotential2014}), hereafter MW-only potential, which was fitted to a large variety of data on the MW, serving both as a simple and accurate model for our galaxy's potential. This potential contains three components consisting of a stellar bulge, a disk, and a dark matter halo, with a total mass of $0.8\times10^{12}M_\odot$. In Appendix~\ref{app:mwmass} we check how our predictions are affected when we consider a MW mass of $1.6\times10^{12}M_\odot$. We considered scenarios with and without the inclusion of the LMC potential, whereas, for the latter case, we used the potential defined by \citet{vasiliev_tango_2021}, using \textsc{agama} \citep{vasiliev_agama_2019}, with a MW mass of $0.88\times10^{12}M_\odot$ and a LMC mass of $1.5\times10^{11}$M$_\odot$. \par
If the Crater-Leo objects were accreted in one common group, then the predicted proper motions should match the measurements of the proper motions we have available today. 

\subsubsection{Intrinsic scatter}\label{sec:mcmc}
To account for the intrinsic scatter in the specific energy $E$ and angular momentum $h$ of the group that are is considered in the Lynden-Bell method, we adopted a Markov Chain Monte Carlo (MCMC) approach. Assuming that the uncertainties are Gaussian and that the measurements are independent, the likelihood of a model $y=y(x)$ is given, generically, by
\begin{equation}
\mathcal{L} = \prod_i \frac{1}{\sqrt{2\pi\sigma^2}}\exp\left[-\frac{(y_i-y(x_i))^2}{2\sigma^2}\right]
\label{eq:generic-likelihood}
\end{equation}
Taking into account our model, given by Eq.~\ref{eq:Er}, and assuming that our uncertainties are underestimated by an intrinsic scatter, $\sigma_\mathrm{is}$, the uncertainty for each measurement $n$ becomes
\begin{equation}
\epsilon_n^2 = \sigma_\mathrm{obs}^2+\sigma^2_\mathrm{is}.
\label{eq:sigma_total}
\end{equation}
Our likelihood function will then be
\begin{equation}
\ln{\mathcal{L}}(y|x,\sigma,h,E,f) = -\frac{1}{2}\sum_n\left[\frac{(y_n-[E-\frac{1}{2}h^2x_n^{-2}])^2}{\epsilon_n^2}+\ln{(2\pi \epsilon_n^2)}\right]
\label{eq:likelihood}
\end{equation}
To determine the posterior probability function of $E$ and $h$ given by
\begin{equation}
p(h,E,f|x,y,\sigma) \propto p(h,E,f)\ \mathcal{L}
\label{eq:posterior}
\end{equation}
with $\mathcal{L}$ being our likelihood function, we need the prior function $p(h,E,f)$. For this, we use $0.5h_0 < h < 1.5h_0$, $0.5E_0 < E < 1.5E_0$ and $0.5\sigma_\mathrm{is_0} < \sigma_\mathrm{is} < 1.5\sigma_\mathrm{is_0}$, with the $h_0, E_0$ being determined by the fit described in the previous Sect.~\ref{sec:pudim}, and $\sigma_\mathrm{is_0}$ being determined by the maximum likelihood result of this model. These are the range of values that allow us to include intrinsic scatter consistent with how far in specific energy the objects are from the best fit, while still making sure that the objects belong to the same group. \par 
After this setup, we sample this distribution using \texttt{emcee} \citep{emcee}. We start by initializing the walkers ($n=250$) around the maximum likelihood result and then run $10^4$ steps of MCMC so they can explore the full posterior distribution. The first half of the steps generated were discarded as a conservative burn-in criterion.

\color{black}

\subsubsection{Looking for additional members}\label{sec:pudim-pt2}
In addition to the already considered group members, other dwarf galaxies or globular clusters around the Milky Way could in principle be associated to the group as well. To further explore other possible members of this association, we once again build upon the Lynden-Bell method. At the time, \cite{lynden-bell_ghostly_1995} did not have 3D positions and velocities available and, for that reason, it was not possible to determine the orbital poles of the objects. To overcome this, they used the Galactocentric vectors of every object and assumed that if two of them (with $\Vec{a}$ and $\Vec{b}$, respectively) were part of the same group, their common orbital pole would lie along $P = \pm \Vec{a}\times \Vec{b}$. The same would happen if several objects (with, for example, $\Vec{c}$ and $\Vec{d}$) are members of the same group: $\pm \Vec{a}\times \Vec{d}$, $\pm \Vec{b}\times \Vec{d}$ and $\pm \Vec{c}\times \Vec{d}$ would have directions close to $P$. Instead of determining orbital poles, they determine putative pole positions by pairing every object with every other. \par

To find additional members, we take the proposed group and follow the method previously described. We start by checking which objects in our data set have putative pole positions closer than $5\degree$ to the ones of our group, i.e. which objects can lie along the great circle that passes close to the group. This is a generous value since all objects of the putative group are closer than $1\degree$. We keep the objects that pass this first criterion since we consider that they can share similar specific angular momentum with the rest of the members. \par 
Following the discovery that satellites are usually accreted in 3-5 members by \citet{li_infall_2008}, we create new groups with the proposed Crater-Leo group plus an additional member that passed the first criterion. We then apply a second criterion to check whether these new objects are likely to share a specific angular momentum and specific energy with our original group or if their orbital poles are just close, without being related. To do this, we use the fitting method previously described and discard the groups that have relative errors in $E_r$ and $h_0$ $>10\%$ since objects accreted together should lie on a line. The new groups that pass this test are considered likely groups and their proper motions are predicted. \par 
As a result of \textit{Gaia} DR3 and HST, we currently have proper motions not only for the most massive dwarfs of the Milky Way but also for some of its faintest satellites (see e.g. \citealp{Fritz2018_gaiadr2}). Thus, in the last step, the predicted proper motions are compared with the measured proper motions, and if the predictions match the observations within $2\sigma$, the group of objects deserves further inspection.


\section{Results}\label{sec:results}
By using the Monte Carlo realisations described in Sect.~\ref{sec:methods}, we estimate the orbital poles for the Crater-Leo objects individually and were able to get an idea if the objects can share specific angular momentum. In Sect.~\ref{sec:res-orbitalpoles}, we present them and discuss their implications. Using the Lynden-Bell method, discussed in Sect.~\ref{sec:pudim}, we predict the proper motions of the Crater-Leo objects. To do this, we use the properties of Crater 1, Crater II, Leo II, Leo IV, and Leo V, discussed in Sect.~\ref{sec:group}. Here, in Sect.~\ref{sec:predicted-pm}, we present the predictions that we obtain for the proposed group. Due to lack of agreement between the measurements and the predictions for Crater II, and the fact that this object does not share a common orbital pole with the others as discussed in Sect.~\ref{sec:res-orbitalpoles}, we repeat the analysis without this object in Sect.~\ref{sub:remove-craterii}. We also check in Sect.~\ref{sub:additional-members} if any other satellites can be part of this putative group. Next, we determine the orbital properties of the objects, described in Table~\ref{tab:tabela2}, and analyse their orbital histories in Sect.~\ref{sec:orbital-parameters}. Finally, we check if the LMC infall had an impact on the current measured orbital poles in Sect.~\ref{sec:effect-lmc}. We predict the proper motions of the objects taking into account the shift in velocities and positions that they would have had if the dark matter halo of the MW was not affected by the nearby LMC.

\subsection{Orbital poles}\label{sec:res-orbitalpoles}
\begin{figure*}
    \includegraphics[scale=0.5]{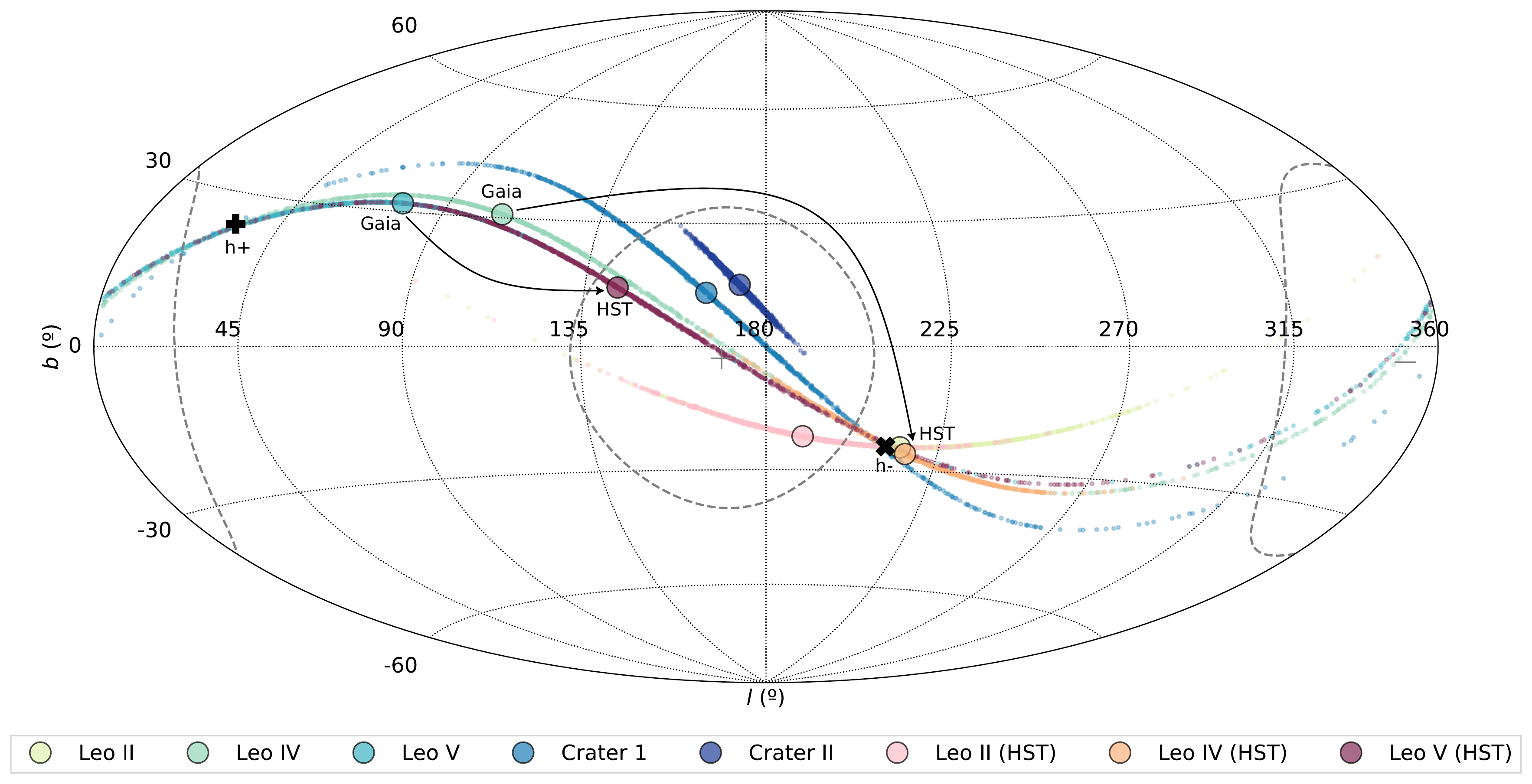}
    \caption{Most likely orbital poles (directions of angular momenta) represented by the large circles and their uncertainties represented by small circles in the corresponding colours for the proposed Crater-Leo objects, based on the proper motions discussed in Sect.~\ref{subsec:propermotions}. The uncertainties in orbital pole directions are obtained by Monte Carlo sampling of the proper motion and distance uncertainties (1000 realisations). The assumed VPOS pole is represented by the grey "+" and the opposite normal direction by "-", with the black circles containing $10\%$ of the sky around them. The proposed group poles are represented by the bold black "x" for the negative circulation, $h-$, and by "+" for the positive circulation $h+$.}
    \label{fig:orbital_poles}
\end{figure*}

The orbital poles that we derive for the Crater-Leo objects using their measured proper motions and for the proposed group are shown in Figure~\ref{fig:orbital_poles}. It is possible to see that all objects are consistent with co-orbiting the VPOS.

From this figure, one can directly identify that Crater II possibly does not belong to a common group of satellite galaxies. However, the other four objects seem to intercept within their uncertainties. It is also possible to notice that using the new measurements from HST of the proper motions of Leo IV and Leo V, a common orbital pole between the possible associated objects becomes more likely. These new HST proper motions also position the most likely orbital pole closer to the VPOS of the MW, as it is possible to see in this figure. 

\subsection{Predicted proper motions}\label{sec:predicted-pm}
\begin{figure}[ht]
    \centering
    \includegraphics[width=\columnwidth]{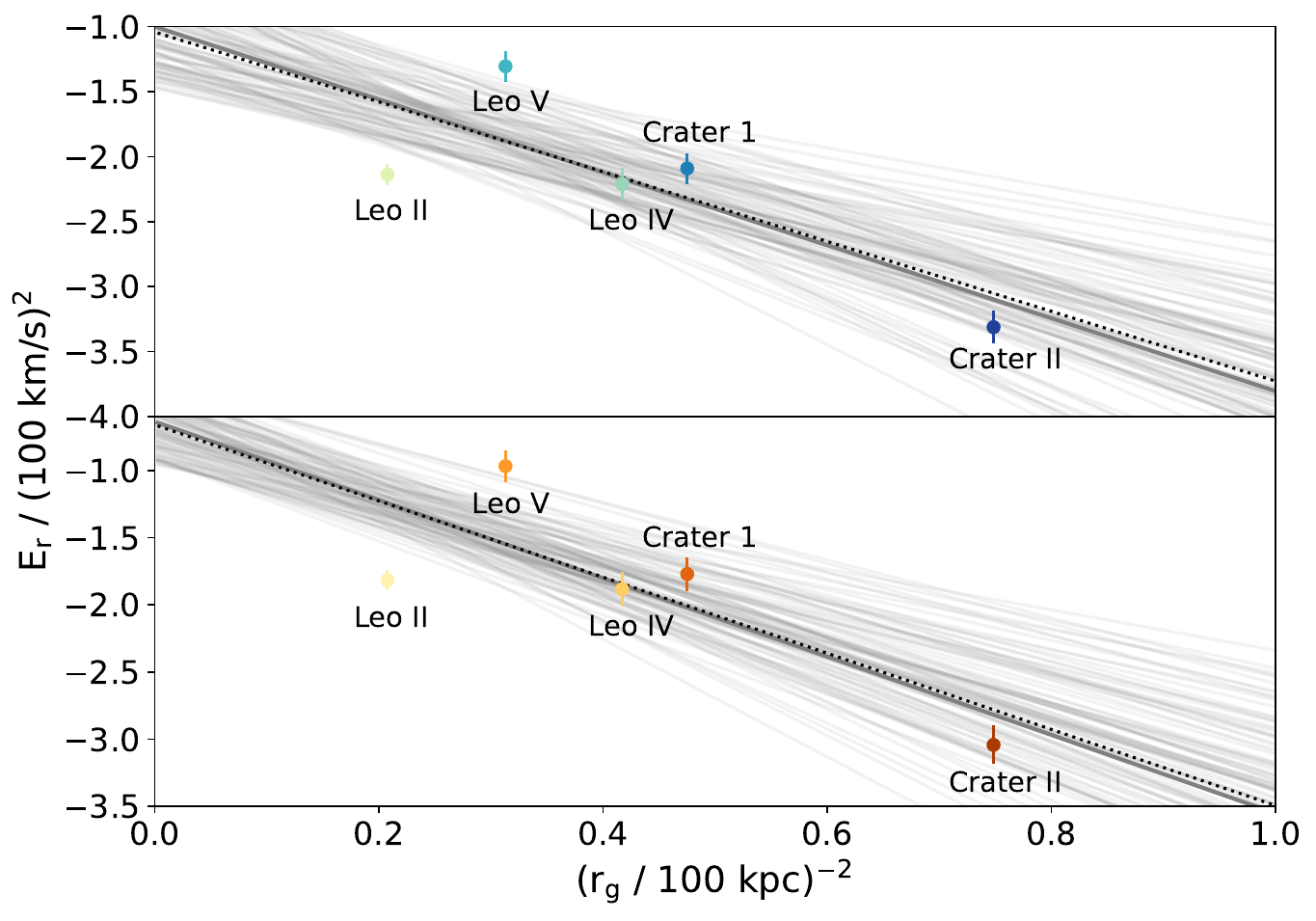}
    \caption{$E_\mathrm{r}$ plotted against $r^{-2}$ for the proposed group. The coloured points represent the measured quantities and the associated uncertainties for each object. The grey solid line shows the fit that we get from the Lynden-Bell method by assuming that the objects under study share $E$ and $h$. The dashed lines represent the mean fit obtained from the MCMC method, with the thin solid lines representing 100 random realisations of it. \textbf{Top:} The fit using MW-only potential. \textbf{Bottom:} The fit using the MW+LMC potential.}
    \label{fig:fit}
\end{figure}

Adopting the method previously discussed in Sect.~\ref{sec:pudim}, we can determine the specific angular momentum and energy for the group. Objects that were accreted together should share these quantities, and, consequently, align in a straight line when we plot $E_r$ against $r^{-2}$, allowing us to determine their values for the group. This plot is represented in Figure~\ref{fig:fit}, for both potentials. As one can see, all the objects lie close to the fitted line. Using the Lynden-Bell method, we get a specific angular momentum of $h_0 = 2.37\times 10^4$ kpc km s$^{-1}$ and specific energy of $E_0 = -0.99\times10^{4}$ km$^{2}$s$^{-2}$ for the considered group. When we perform the MCMC method described in Sect.~\ref{sec:mcmc} we get a specific angular momentum of $h_0 = 2.32^{+0.29}_{-0.32} \times 10^4$ kpc km s$^{-1}$ and specific energy of $E_0 = -1.05^{+0.32}_{-0.29} \times10^{4}$ km$^{2}$s$^{-2}$ for the considered group. Including the influence of the LMC, these values change to $h_0 = 2.41\times 10^4$ kpc km s$^{-1}$ and $E_0 = -0.64\times10^4$  km$^{2}$s$^{-2}$. The MCMC method gives $h_0 = 2.38^{+0.25}_{-0.27} \times 10^4$ kpc km s$^{-1}$ and $E_0 = -0.66^{+0.22}_{-0.20}\times10^4$  km$^{2}$s$^{-2}$ for this potential. The values that we get for the group for both the specific angular momentum and the specific energy are similar for both potentials, so it is expected that the proper motions that we predict from here are similar as well. The corner plots of the posterior distributions for the energy and angular momentum for both potentials can be seen in Figure~\ref{fig:cornerplots-group} of the Appendix~\ref{app:mcmc}. 

\begin{figure}[ht]
    \centering
    \includegraphics[width=\columnwidth]{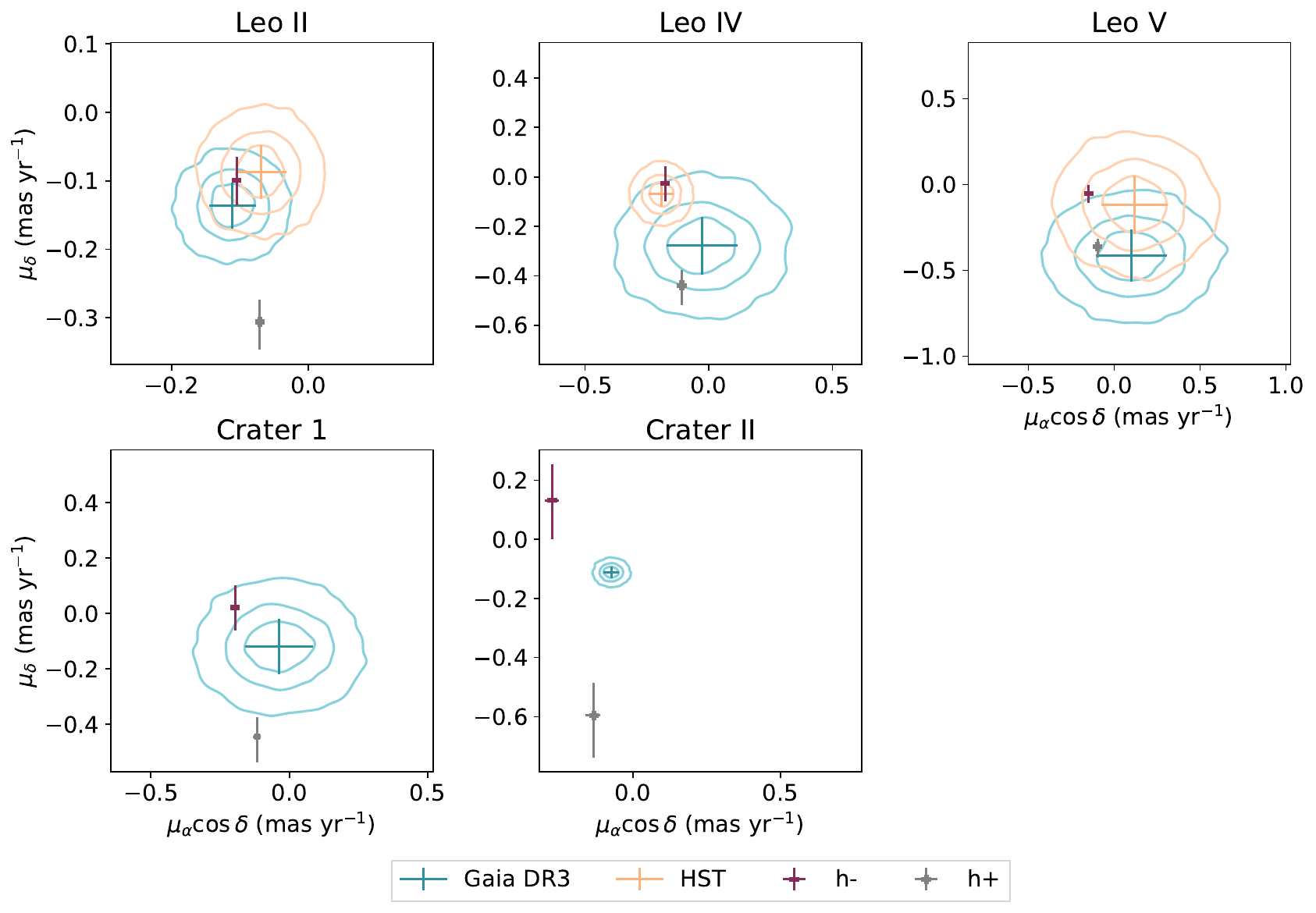}
    \caption{Predicted proper motions of the Crater-Leo objects. The grey "+" and magenta "-" represent the predictions obtained with the negative $h+$ and positive $h-$ circulation about the northern pole, respectively. The blue crosses represent the measured proper motions from \textit{Gaia} DR3 with their uncertainties and the orange crosses represent the proper motion measurements from HST. The corresponding contours correspond to the 34\%, 68\% and 95\% confidence levels for the 2D distribution of the proper motion parameter space.}
    \label{fig:predicted-pm}
\end{figure}

The predicted proper motions that we get for these objects are represented in Figure~\ref{fig:predicted-pm}. The predictions remain nearly identical when we include the influence of the LMC in the potential of the MW. For that reason, they are not shown in the plot. To determine the parameter space of possible values that the measured proper motions can take within their uncertainties, we once again perform Monte Carlo realisations. We also take into account the systematic errors associated with the \textit{Gaia} DR3 proper motion measurements, described in~\cite{Battaglia2022_gaiadr3}.

The predicted proper motions seem to better match the measured ones for Leo II and Crater 1 when we consider negative $h-$ circulation. When we look at Leo IV and Leo V, we also notice that those predictions are closer to the proper motions measured by HST. That is expected since their orbital poles lie close to the rest of the group when we consider the HST measurements. However, the predictions for Crater II do not match the measurements at all and the measured proper motions are well constrained. The possible orbital poles of this satellite suggest that this object cannot share the same angular momentum with the rest of the potential group, and this mismatch of predicted and measured proper motions is another indicator that this galaxy is not associated with it.

\subsubsection{Removing Crater II}\label{sub:remove-craterii}
By looking at the orbital poles of the proposed group, Crater II does not seem to share a similar direction of angular momentum with the other objects. Furthermore, when we look at the line-of-sight velocities of these satellites, Crater II is the most offset one from the common value shared by the other objects (right panel of Figure~\ref{fig:positions}). Additionally, as previously mentioned, this object only shows signs of star formation events at 10.5 Gyr and 12.5 Gyr ago, with no presence of intermediate-age or younger stars \citep{walker_decam_2019}, contrary to the rest of the group. This suggests that this dwarf galaxy is unlikely to have been accreted with the other satellites. Since the predictions are based on the properties of the full ensemble of considered group members, the predicted proper motions will differ if one object is excluded. In order to see if the predicted proper motions better match the measured ones when this object is removed, we repeat the analysis previously described without Crater II. 

\begin{figure}[ht]
    \centering
    \includegraphics[width=\columnwidth]{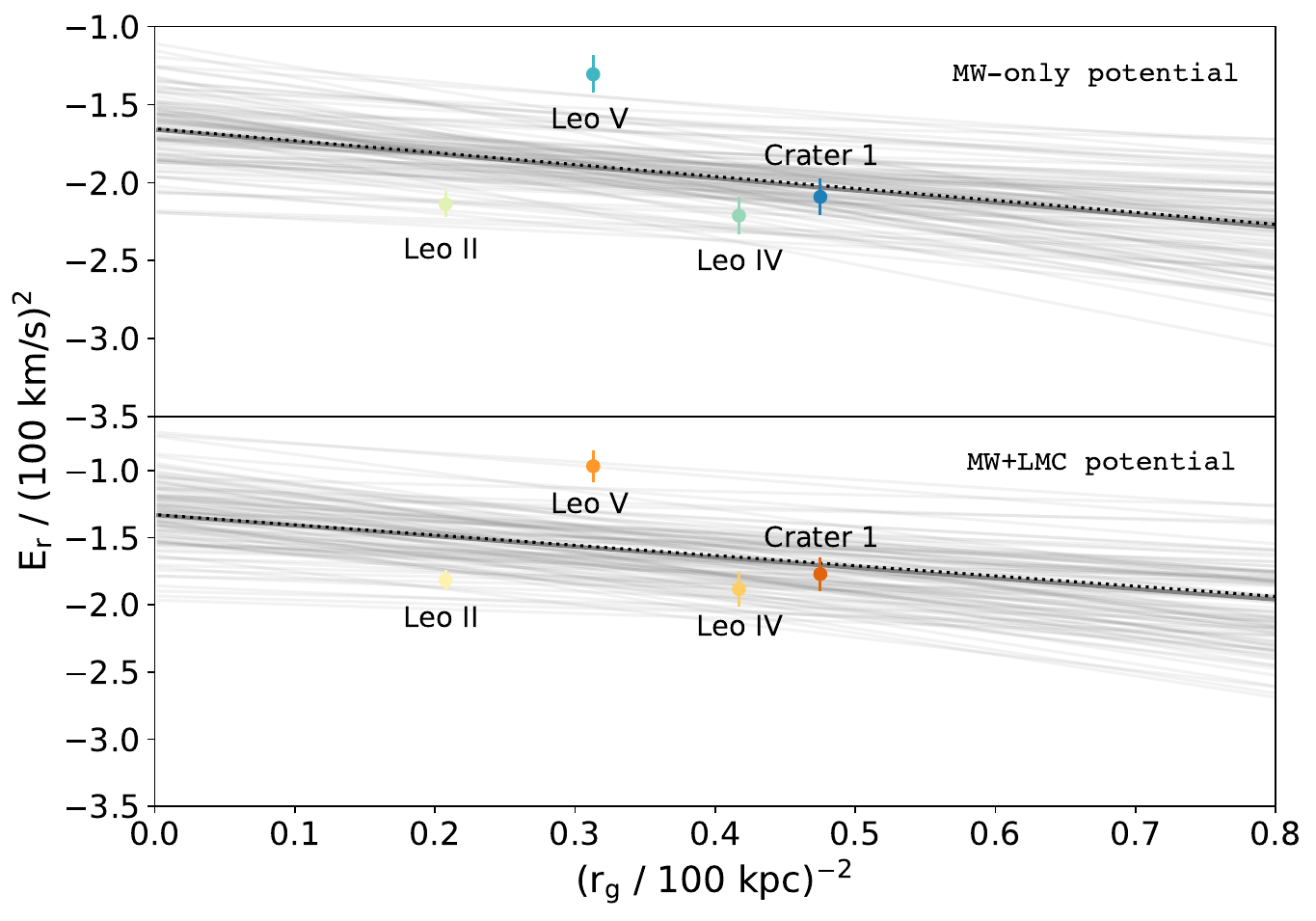}
    \caption{$E_\mathrm{r}$ plotted against $r^{-2}$ after removing Crater II from the group. The coloured points represent the measured quantities and the associated uncertainties for each object. The grey solid line shows the fit that we get from the Lynden-Bell method by assuming that the objects under study share $E$ and $h$. The dashed lines represent the mean fit obtained from the MCMC method, with the thin solid lines representing 100 random realisations of it. \textbf{Top:} The fit using MW-only potential potential. \textbf{Bottom:} The fit using the MW+LMC potential.}
    \label{fig:fit_without_craterii}
\end{figure}

In this case, we get a specific angular momentum of $h_0 = 1.25\times 10^4$ kpc km s$^{-1}$ and specific energy of $E_0 = -1.66\times10^4$  km$^{2}$s$^{-2}$ for the Crater-Leo objects after removing Crater II from the putative group. The MCMC methods gives $h_0 = 1.24^{+0.42}_{-0.42} \times 10^4$ kpc km s$^{-1}$ and $E_0 = -1.65^{+0.26}_{-0.24} \times10^{4}$ km$^{2}$s$^{-2}$. When we use the potential that includes the influence of the LMC, these values change to $h_0 = 1.25\times 10^4$ kpc km s$^{-1}$ and $E_0 = -1.33\times10^4$  km$^{2}$s$^{-2}$. Considering the MCMC method, we get $h_0 = 1.23 ^{+0.42}_{-0.41} \times 10^4$ kpc km s$^{-1}$ and $E_0 = -1.33^{+0.26}_{-0.24}\times10^{4}$ km$^{2}$s$^{-2}$. Both the specific angular momentum and specific energy decrease for the group when we remove this satellite. These fits are represented in Figure~\ref{fig:fit_without_craterii} and the posterior distributions in Figure~\ref{fig:cornerplots-group-without-craterii} of the Appendix~\ref{app:mcmc}. As one can see, the specific angular momentum is not well constrained for this case.

\begin{figure}[ht]
    \centering
    \includegraphics[width=\columnwidth]{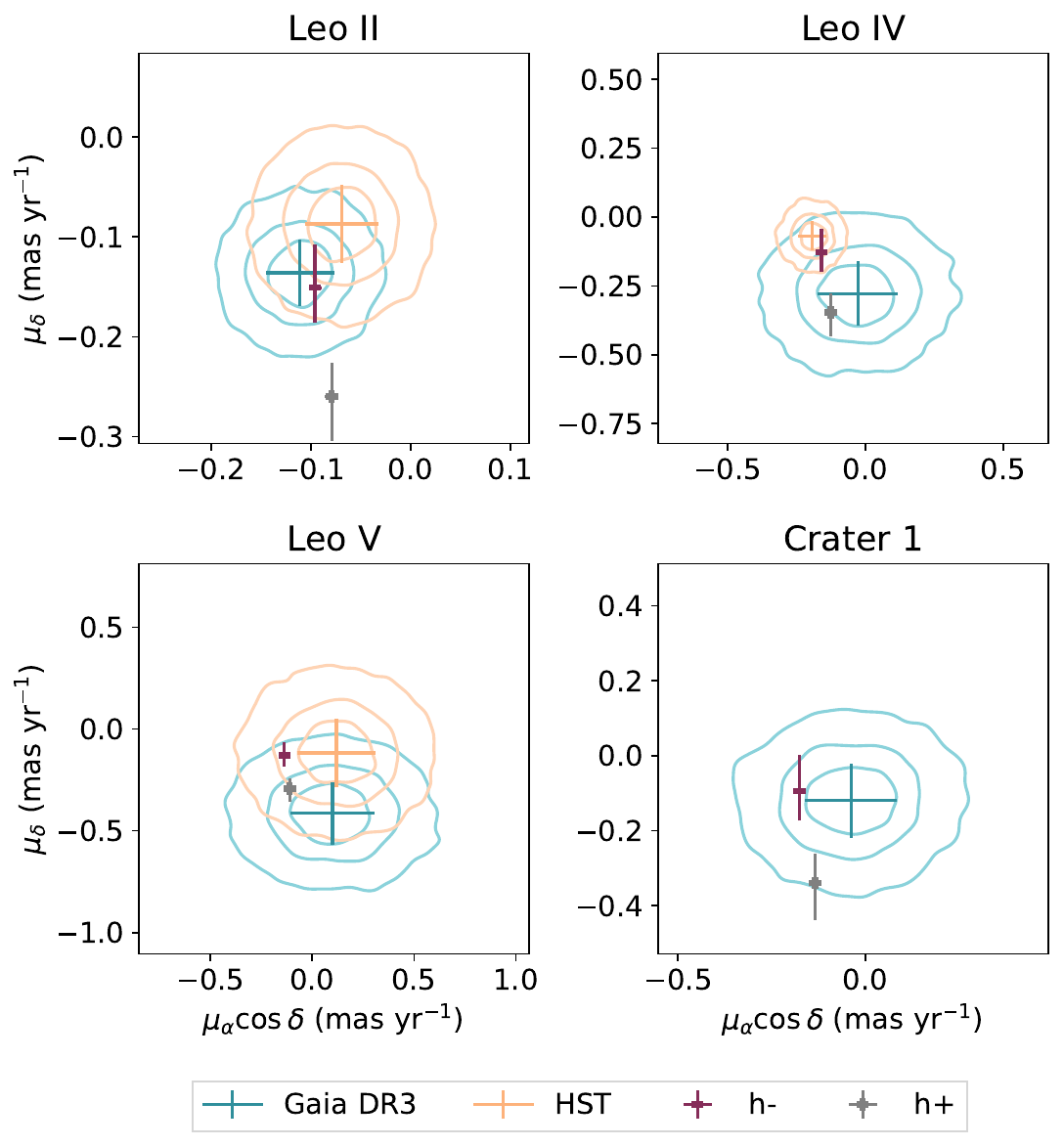}
    \caption{Predicted proper motions of the Crater-Leo objects without Crater II. The grey "+" and magenta "-" represent the predictions obtained with the negative $h+$ and positive $h-$ circulation about the northern pole, respectively. The blue crosses represent the measured proper motions from \textit{Gaia} DR3 with their uncertainties and the orange crosses represent the proper motion measurements from HST. The corresponding contours correspond to the 34\%, 68\% and 95\% confidence levels for the 2D distribution of the proper motion parameter space.}
    \label{fig:predicted-pm-withoutcrater}
\end{figure}

The proper motions that we get when we remove Crater II can be seen in Figure~\ref{fig:predicted-pm-withoutcrater}. The negative $h-$ circulation seems to predict pretty well the proper motion of the remaining objects, being a strong indicator that these objects were accreted together. When we consider the new HST proper motions for Leo IV and Leo V, all the predictions fall within $1\sigma$ confidence level of the measured uncertainties. For Leo II, the prediction falls within the $0.5\sigma$ confidence level. For these reasons, we will perform the rest of the analysis without Crater II.

\subsection{Additional members}\label{sub:additional-members}
To determine which other MW halo objects could be possible additional members of the proposed group, we used a combination of the 3D positions and velocities of dwarf galaxies from \citet{Battaglia2022_gaiadr3} and globular clusters from \citet{baumgardt_globular_2019}. 
We start by estimating the orbital poles of all the objects with distances $r>20$ kpc, where our potential is well approximated by a spherical one. We use the method described in Sect.~\ref{sec:pudim-pt2} and try to find groups with the previously discussed objects (Leo II, Leo IV, Leo V, and Crater 1) and possible additions. \par 

We find five satellite galaxies that match our two first criteria (putative pole positions closer than $5\degree$ to the rest of the group and relative errors smaller than $10\%$ in $E_r$ and $h_0$ when we use the fitting method for the group): Phoenix II, Tucana II \citep{koposov_beasts_2015}, Tucana IV, Tucana V \citep{drlica-wagner_eight_2015}, and Ursa Major II \citep{zucker_curious_2006}. This means that these objects can share similar specific angular momenta and energy with the rest of the proposed group. However, when we compare the predicted proper motions of the possible additional objects with the observed ones from \textit{Gaia} DR3 \citep{Battaglia2022_gaiadr3}, they do not match within $1.5\sigma$. Even increasing this value to $>3\sigma$, the proper motions do not match, so it is very unlikely that they were accreted with the initially proposed group. The predicted proper motions of these objects can be seen in Figure~\ref{fig:predicted-pm-additional-members} of the Appendix. As noticeable from this figure, the proper motions for these objects are not well constrained.


\subsection{Orbital dynamics}\label{sec:orbital-parameters}
\begin{figure*}[ht]
    \centering
    \includegraphics[scale=0.4]{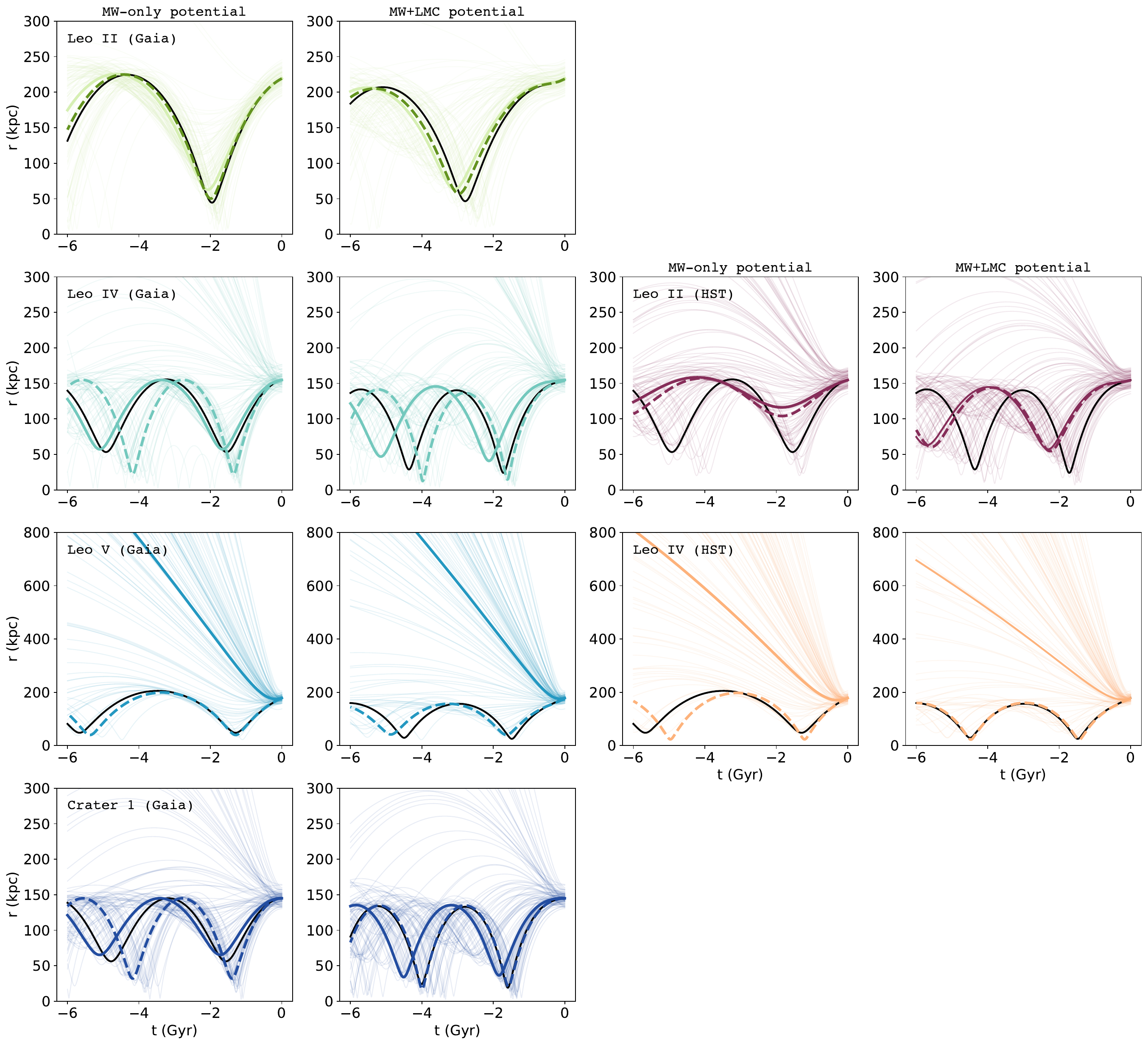}
    \caption{Orbital evolution for the proposed group integrating backwards in time for 6 Gyr using both the MW-only potential and the MW+LMC potential. The thick lines indicate their most likely orbit from the measured proper motions and the thinner lines represent 100 Monte-Carlo realisations within their uncertainties. The black solid lines represent the orbital evolution obtained using the predicted proper motions for each object. The dashed lines represent the mean orbital evolution obtained using only the bound orbits.}
    \label{fig:orbits-MC} 
\end{figure*}

\begin{figure*}
    \centering
    \includegraphics[scale=0.43]{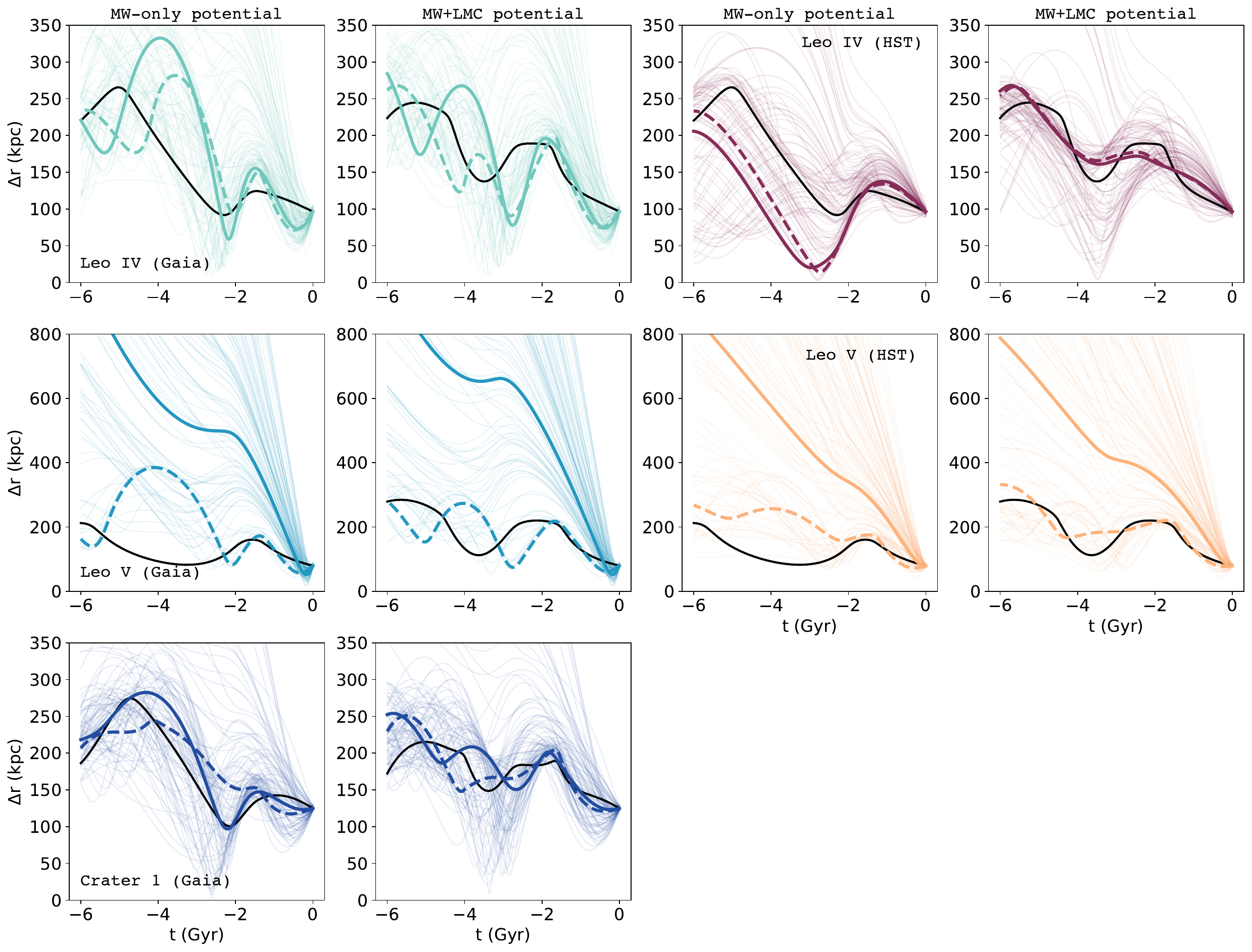}
    \caption{Orbital evolution for the proposed group integrating backwards in time for 6 Gyr for the mutual separation of the objects in relation to Leo II using both the MW-only potential and the MW+LMC potential. The thick lines indicate their most likely orbit from the measured proper motions and the thinner lines represent 100 Monte-Carlo realisations within their uncertainties. The black solid lines represent the orbital evolution obtained using the predicted proper motions for each object.  The dashed lines represent the mean orbital evolution obtained using only the bound orbits.}
    \label{fig:leoii-separation-mc} 
\end{figure*}

For given proper motions (either measured or predicted), we can derive the orbital histories for each object to assess whether they could have been a bound group. We do this by integrating their orbits in the potentials discussed before for the measured proper motions (described in Tables~\ref{tab:proper-motions} and~\ref{tab:hst-proper-motions}), both in relation to the centre of the MW (in Figure~\ref{fig:orbits-MC}) and for the mutual separation in relation to the most massive object, Leo II (in Figure~\ref{fig:leoii-separation-mc}). 

Figure~\ref{fig:orbits-MC} shows the orbital evolution of the Crater-Leo objects for both the MW-only potential and the MW+LMC potential. We include the possible errors within the observational uncertainties, drawn from Monte Carlo realisations. We also plot in black the orbits derived from the predicted proper motions.
It is possible to see that all objects reached their apocentre at the present time. For the MW-only potential, all objects except Leo V reached their pericentre around the same time ($\sim$ 2 Gyr ago) and with similar values ($\sim$ 60 kpc). When we consider the HST measurement for Leo IV, its orbit becomes almost circular, so its pericentre is further away from the MW centre. For Leo II and Crater 1, the predicted orbit is similar to the most likely orbit from the measurements. For the remaining objects, even though the most likely and the predicted orbits do not match, the predicted orbit is still consistent within the uncertainties. When we look only at the bound orbits of Leo IV and Leo V, this consistency is even larger, suggesting that with future proper motions we may get measured orbits closer to the predicted ones. Figure~\ref{fig:face-on-mc} of the Appendix shows the same thing but for the face-on view of the orbits. \par

When we include the LMC in the potential, the pericentres of the satellites no longer happen at the same time, but rather one at a time, with an increased time with decreasing distance to the MW. The proper motion of Leo V is still too unconstrained to reach such conclusions since this object appears to be on first infall into the MW throughout its orbit history when we consider the most likely PM. If this was the real orbit of this dwarf galaxy, then it would mean that this object is just passing close to the MW at the present time, and by chance, ended up close to the other objects. The orbital parameters that we get for this galaxy also show this, since we get unreasonable values for the parameters and extremely large errors. However, when we look at the possible orbits within the observational errors, some orbits are bound, for both the \textit{Gaia} and the HST proper motions. Considering only those, Leo V is also at its apocentre at the present time. When we look back in time, its pericentre matches the time when the other objects also reached their pericentres.  \par

Figure~\ref{fig:leoii-separation-mc} shows the separation of Leo IV, Leo V and Crater 1 with respect to Leo II, for the MW-only potential and the MW+LMC potential. We use the \textit{Gaia} DR3 proper motion to get the orbit of Leo II since it is the closest to the prediction. This helps us to understand how close the other objects have been to Leo II during their orbital history. In the MW-potential, if we consider the HST proper motion for Leo IV, this object gets as close as $\approx 20$ kpc to Leo II $~\sim 3$ Gyr ago. However, if we consider the \textit{Gaia} proper motion, both Leo IV and Crater 1 are closer to Leo II $\sim 2$ Gyr ago for the MW-only potential, and $\sim 2$ Gyr ago for the MW+LMC potential. However, all of them are close to Leo II at the present time. This time, the predicted orbit is quite similar for Crater 1 when we consider the MW-only potential. Looking at Leo IV, the prediction follows the behaviour obtained with both potentials. But, for Leo V, once again, most orbits are unbound due to the high uncertainties of the proper motions. Yet, looking only at the bound orbits, they also follow the same behaviour as the prediction, especially for the MW+LMC potential, being another indicator that these objects were indeed accreted as a group. Our preferred orbits, indicated by the black lines, have the galaxies at comparable distances from each other some $\sim 3$ Gyr ago as they are today. This is consistent with our finding that there is a physical association between these galaxies, and that they are not just a chance alignment at the present time. Further back in time, our orbit calculations show a trend of increasing distances. This is opposite to what may be expected (namely, that the galaxies were closer together in the past, and have suffered subsequent diffusion and phase mixing). However, this could easily be due to shortcomings in our orbit calculations, which, e.g. do not account for dynamical friction from the MW, or to hierarchical growth of the MW mass over cosmological timescales. Furthermore, the mutual gravity of the objects was not considered, and since Leo II is more massive than the rest of the objects, it can have an effect on their proximity when we look at their mutual separation.

In both Figures~\ref{fig:orbits-MC} and ~\ref{fig:leoii-separation-mc}, it is possible to see that for all objects, there are orbits within the observational uncertainties that make no physical sense for bound satellites, for the same reasons previously stated for Leo V. Evidently, the subset of allowed orbits is smaller for the objects with larger errors. When we compare only the bound orbits with the orbits derived from the predicted PMs, they are similar. This means that the physically allowed orbits for our objects are consistent with the ones that we get when we assume that they share similar specific angular energy and angular momentum. Besides being consistent with the observations, our predictions also provide a physical explanation for the coherence among the Crater-Leo objects. \par

We can now look at the derived orbital parameters (i.e. pericentre, $r_{\text{peri}}$, apocentre, $r_{\text{apo}}$, eccentricity, $e$, and time since the last pericentre, $t_{\text{peri}}$) of the Crater-Leo objects. To determine the errors in the parameters, we draw $10^4$ Monte Carlo samples as before. We use the coordinates, distances and velocities described in Table~\ref{tab:properties} and their corresponding errors for this purpose. The results that we get are summarized in Table~\ref{tab:tabela2}. For Leo IV and Leo V, we also compare the derived properties that we get using our HST proper motions. Since some orbits that we get within the uncertainties are unbound (as can be seen in Figure~\ref{fig:orbits-MC}), we only consider the bound orbits to perform this calculation. The unbound orbits are likely not physical, but instead due to the very uncertain nature of the proper motion measurements of the objects. Furthermore, since the most likely orbit of Leo V -- for both the HST and \textit{Gaia} measurements -- is unbound, we consider the mean value of the MC realisations as the central value instead. For consistency, we do this for all objects. Once again, we only derive the orbital properties using the predicted proper motions assuming $h-$\, since a common group accretion requires a common orbiting direction and the objects prefer this one. For the latter, the errors were estimated taking into account the error in all parameters except the proper motions. \par

\begin{sidewaystable}
\centering
\caption{Derived orbital parameters of Crater-Leo objects: the pericentre ($r_{\text{peri}}$), apocentre ($r_{\text{apo}}$), eccentricity ($e$). We compare the derived orbital parameters by using the proper motions described in Table~\ref{tab:proper-motions} (\textit{Gaia} DR3) with the ones derived by using the predicted proper motions with the Lynden-Bell method described in Sect.~\ref{sec:predicted-pm} (Predicted). For Leo IV and Leo V, we also derive the orbital parameters by using our new HST proper motions (HST). For all the derived orbital parameters, the properties described in Table~\ref{tab:properties} were used. We also mention the percentage of bound orbits that we get for each object.} We do this for both the MW-only potential and the MW+LMC potential.

\begin{tabular}{lccccccccccccccccc}
\hline
\textbf{MW-only} & \multicolumn{3}{c}{$r_{\text{peri}}$ (kpc)}  & \multicolumn{3}{c}{$r_{\text{apo}}$ (kpc)}  & \multicolumn{3}{c}{$e$}  & \multicolumn{3}{c}{$t_{\text{peri}}$ (Gyr)}  & \multicolumn{2}{c}{{Bound orbits (\%)}}   \\
\multicolumn{1}{l}{Object} & \textit{Gaia} & HST & \multicolumn{1}{l}{Predicted} & \textit{Gaia} & HST & \multicolumn{1}{l}{Predicted} & \textit{Gaia} & HST & Predicted & {\textit{Gaia}} & {HST} & {Predicted} & {\textit{Gaia}} & {HST} \\ \hline
Leo II    &  $71^{+26}_{-18}$   & -                 & $48^{+16}_{-10}$ &  $226^{+9}_{-9}$    & -                   & $225^{+7}_{-7}$  &  $0.55^{+0.11}_{-0.10}$ & -                      & $0.66^{+0.07}_{-0.08}$ & $-2.1^{+0.1}_{-0.1}$ & -                    & $-1.9^{+0.1}_{-0.1}$ & 83\% & - \\
Leo IV    &  $87^{+24}_{-24}$   & $110^{+14}_{-25}$ & $56^{+16}_{-10}$ &  $176^{+60}_{-10}$  & $185^{+54}_{-10}$   & $156^{+6}_{-5}$  &  $0.39^{+0.16}_{-0.11}$ & $0.27^{+0.13}_{-0.06}$ & $0.48^{+0.08}_{-0.09}$ & $-1.3^{+0.5}_{-0.4}$ & $-1.2^{+0.3}_{-0.2}$ & $-1.6^{+0.1}_{-0.1}$ & 70\% & 91\% \\
Leo V     &  $93^{+17}_{-24}$   & $87^{+18}_{-23}$  & $40^{+13}_{-8}$  &  $236^{+33}_{-16}$  & $234^{+34}_{-15}$   & $207^{+10}_{-8}$ &  $0.47^{+0.13}_{-0.04}$ & $0.49^{+0.13}_{-0.05}$ & $0.62^{+0.06}_{-0.06}$ & $-1.2^{+0.2}_{-0.1}$ & $-1.2^{+0.2}_{-0.1}$ & $-1.3^{+0.1}_{-0.1}$ & 14\% & 19\%\\
Crater 1  &  $76^{+24}_{-20}$   & -                 & $59^{+15}_{-9}$  &  $154^{+53}_{-7}$   & -                   & $145^{+5}_{-4}$  &  $0.39^{+0.16}_{-0.11}$ & -                      & $0.43^{+0.07}_{-0.08}$ & $-1.8^{+0.5}_{-1.3}$ & -                    & $-1.6^{+0.1}_{-0.1}$ & 70\% & - \\
\hline
\textbf{MW+LMC} & \multicolumn{3}{c}{}               & \multicolumn{3}{c}{} & \multicolumn{3}{c}{} & \multicolumn{3}{c}{}  & \multicolumn{2}{c}{} \\ \hline
Leo II    &  $111^{+11}_{-20}$  & -                  & $48^{+18}_{-10}$ &  $217^{+23}_{-10}$ & -                  & $225^{+7}_{-7}$  &  $0.41^{+0.13}_{-0.09}$ & -                      & $0.66^{+0.07}_{-0.08}$ & $-1.8^{+0.1}_{-0.3}$ & -                    & $-2.0^{+0.1}_{-0.1}$ &  77\% & -\\
Leo IV    &  $81^{+27}_{-18}$   & $94^{+8}_{-16}$    & $56^{+16}_{-10}$ &  $163^{+56}_{-10}$ & $163^{+50}_{-10}$  & $156^{+6}_{-5}$  &  $0.42^{+0.14}_{-0.16}$ & $0.34^{+0.14}_{-0.09}$ & $0.48^{+0.08}_{-0.09}$ & $-1.6^{+0.6}_{-1.2}$ & $-1.2^{+0.1}_{-0.3}$ & $-1.6^{+0.1}_{-0.1}$ &  65\% & 97\%\\
Leo V     &  $111^{+18}_{-30}$  & $100^{+23}_{-27}$  & $50^{+13}_{-9}$  &  $198^{+54}_{-14}$ & $184^{+51}_{-12}$  & $207^{+10}_{-8}$ &  $0.32^{+0.18}_{-0.09}$ & $0.35^{+0.17}_{-0.10}$ & $0.62^{+0.06}_{-0.06}$ & $-1.3^{+0.3}_{-0.5}$ & $-1.5^{+0.4}_{-0.4}$ & $-1.3^{+0.1}_{-0.1}$ &  17\% & 26\% \\
Crater 1  &  $52^{+27}_{-14}$   & -                  & $59^{+15}_{-9}$  &  $154^{+26}_{-6}$  & -                  & $145^{+5}_{-4}$  &  $0.54^{+0.12}_{-0.13}$ & -                      & $0.43^{+0.07}_{-0.08}$ & $-2.3^{+0.5}_{-1.0}$ & -                    & $-1.6^{+0.1}_{-0.1}$ &  78\% & - \\
\hline
\label{tab:tabela2}
\end{tabular}
\end{sidewaystable}

As anticipated, taking into account that the predicted PMs for both potentials are nearly identical, the orbital properties obtained using these values are quite similar. \par
We can compare these values with the ones derived by \citet{Fritz2018_gaiadr2} since they use the same MW-only potential. However, they use \textit{Gaia} DR2 to obtain them, and the proper motions of the objects under consideration have larger errors in their data. Their derived parameters are described in Table~\ref{tab:fritz}. Our derived values for both the predicted proper motions and the measured ones match their values within the errors. The proper motions of Leo V, even when considering our new HST ones, are too unconstrained. This biases the derived orbital properties to higher values when we consider the MW-only potential. We can also compare our derived orbital parameters with \citet{Battaglia2022_gaiadr3}, since we use their \textit{Gaia} DR3 PMs. However, we assume a slightly different MW mass and we remove the unbound orbits and that leads to slightly different results. They use a light ($M_\mathrm{MW} = 0.9\times10^{12}M_\odot$) and a heavy potential ($M_\mathrm{MW} = 1.6\times10^{12}M_\odot$) for the MW, and our results are closer to the ones obtained by assuming a lighter potential, since our assumed mass is closer to that one ($M_\mathrm{MW} = 0.8\times10^{12}M_\odot$).

\begin{table}
	\centering
	\caption{Pericentre ($r_{\text{peri}}$), apocentre ($r_{\text{apo}}$), eccentricity ($e$) derived from \textit{Gaia} DR2 proper motions from \citet{Fritz2018_gaiadr2}.}
	\label{tab:fritz}
	\begin{tabular}{lccc} 
		\hline
		Object & $r_{\text{peri}}$ (kpc) & $r_{\text{apo}}$ (kpc) & $e$ \\
		\hline
		Leo II & $67^{+154}_{-52}$ & $248^{+613}_{-26}$ & $0.67^{+0.26}_{-0.39}$   \\
		Leo IV & $153^{+8}_{-87}$ & $26071^{+46619}_{-25908}$ & $0.989^{+0.007}_{-0.56}$   \\
		Leo V  & $168^{+12}_{-104}$ & $27704^{+45671}_{-27495}$ & $0.988^{+0.007}_{-0.46}$   \\
            Crater 1 & $81^{+65}_{-70}$ & $159^{+7932}_{-16}$ & $0.68^{+0.3}_{-0.43}$  \\
		\hline
	\end{tabular}
\end{table}

\subsection{Effect of the LMC on the orbital poles}\label{sec:effect-lmc}
It was argued in \cite{garavito-camargo_clustering_2021} that the infall of a massive LMC can perturb the MW halo, affecting the orbital poles of its satellites. This happens because, as described in \cite{gomez_and_2015}, as the LMC passes through its pericentre, the orbital barycentre of the MW-LMC system shifts away from the centre of the MW's disk. Consequently, the resulting reflex motion of the disk is represented by an all-sky dipole pattern in radial velocities (e.g. \citealp{gomez_and_2015}, \citealp{garavito-camargo_hunting_2019}, and \citealp{petersen_reflex_2020}). \par

In \cite{garavito-camargo_clustering_2021}, they reproduce this behaviour by adopting different reference centres of mass for different radial shells, consequently changing the positions and velocities of the satellites. Similarly to \cite{pawlowski_effect_2022}, we invert the approach of \cite{garavito-camargo_clustering_2021} and subtract the shifts from the observed MW dwarf galaxies created by the LMC influence. Following the method described in \cite{pawlowski_effect_2022}, we reproduce the centres of mass shifts from Figures 4 and 6 of \cite{garavito-camargo_clustering_2021} and subtract the respective distance-dependent values from the Cartesian positions and velocities of the observed MW satellites (previously described in Table~\ref{tab:properties}). This allows us to determine the positions and velocities that the satellites would have had relative to the centre of the MW dark matter halo if the MW at the centre was not affected by the nearby LMC. These new values are described in Table~\ref{tab:lmc_effect_shifts} of the Appendix, as well as a Figure~\ref{fig:lmc-effect-shifts} that shows these shifts. \par 

Here, we repeat the previous analysis to get proper motion predictions taking into account this shift, but this time only with the MW-only potential, since we are hypothesizing that our group fell onto the MW halo before the LMC. \par

\begin{figure}[ht]
    \centering
    \includegraphics[width=\columnwidth]{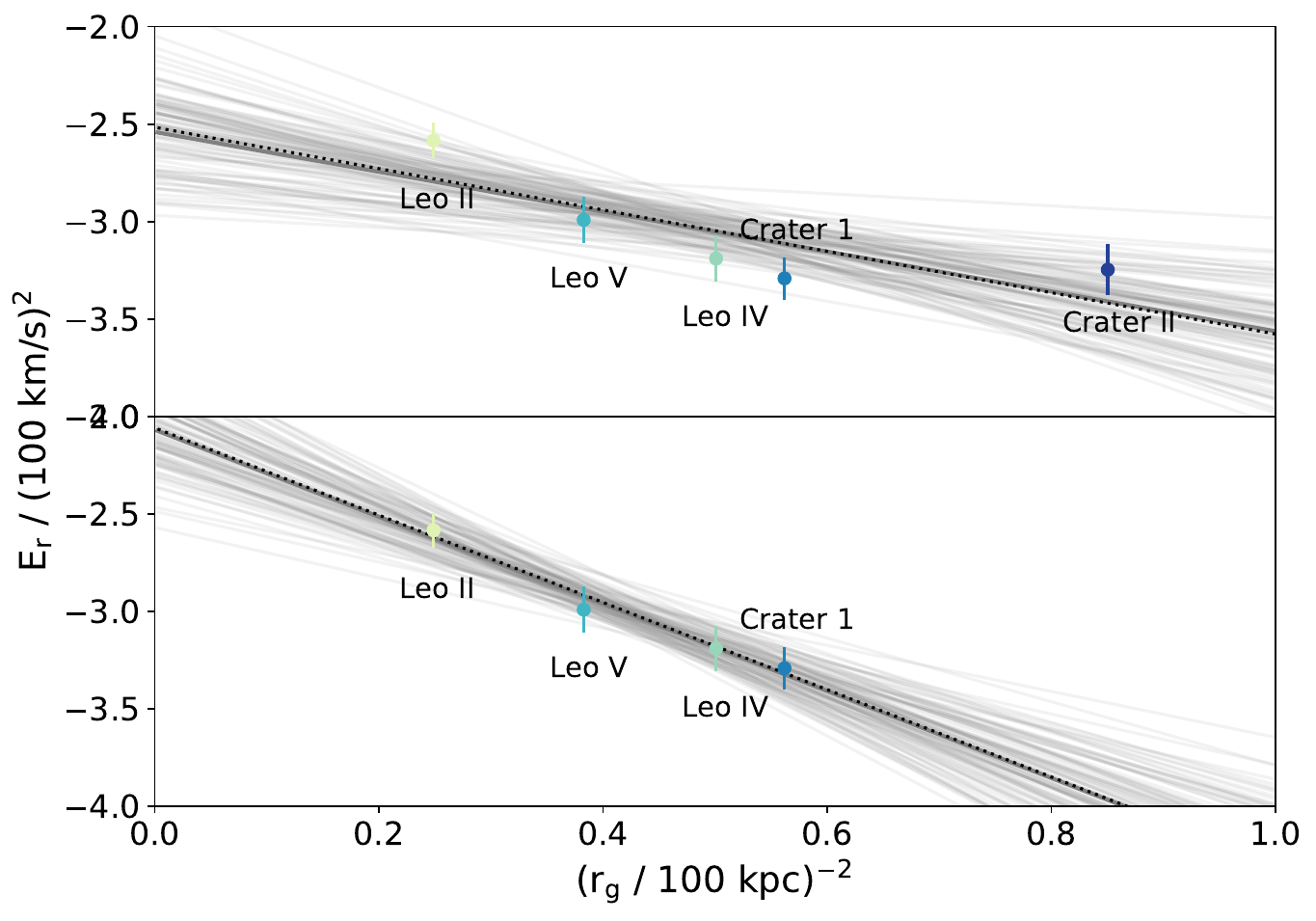}
    \caption{$E_\mathrm{r}$ plotted against $r^{-2}$ after applying the shifts to the positions and velocities from \cite{garavito-camargo_clustering_2021}. The coloured points represent the shifted quantities and the associated uncertainties for each object. The grey solid line shows the fit that we get from the Lynden-Bell method by assuming that the objects under study share $E$ and $h$. The dashed lines represent the mean fit obtained from the MCMC method, with the thin solid lines representing 100 random realisations of it. \textbf{Top:} The fit considering the original proposed group. \textbf{Bottom:} The fit after removing Crater II.}
    \label{fig:fit_lmc_effect}
\end{figure}

Figure~\ref{fig:fit_lmc_effect} shows the new fits that give us $h_0 = 1.43 \times 10^4$kpc km s$^{-1}$ and specific energy of $E_0 = -2.54 \times10^{4}$km$^{2}$s$^{-2}$ when we include Crater II, and $h_0 = 2.11 \times 10^4$kpc km s$^{-1}$ and specific energy of $E_0 = -2.07 \times10^{4}$km$^{2}$s$^{-2}$ when we remove it. Again, applying the MCMC method described in Sect.~\ref{sec:mcmc}, gives  $h_0 = 1.44^{+0.27}_{-0.32} \times 10^4$ kpc km s$^{-1}$, $E_0 = -2.53^{+0.22}_{-0.22} \times10^{4}$ km$^{2}$s$^{-2}$ and $h_0 = 2.12^{+0.19}_{-0.21} \times 10^4$ kpc km s$^{-1}$, $E_0 = -2.06^{+0.18}_{-0.18} \times10^{4}$ km$^{2}$s$^{-2}$, with and without Crater II, respectively. Contrary to what happened before when we removed this object, this time the specific angular momentum and energy increase after removing it. From this figure, it is possible to see that the alignment with a linear fit becomes much better (see Figures~\ref{fig:fit} and \ref{fig:fit_without_craterii}) when we include this effect. The corner plots of the posterior distributions for the energy and angular momentum for both potentials can be seen in Figure~\ref{fig:cornerplots-lmc-effect} of the Appendix~\ref{app:lmceffect}.

\begin{figure}
   \centering
    \includegraphics[width=\columnwidth]{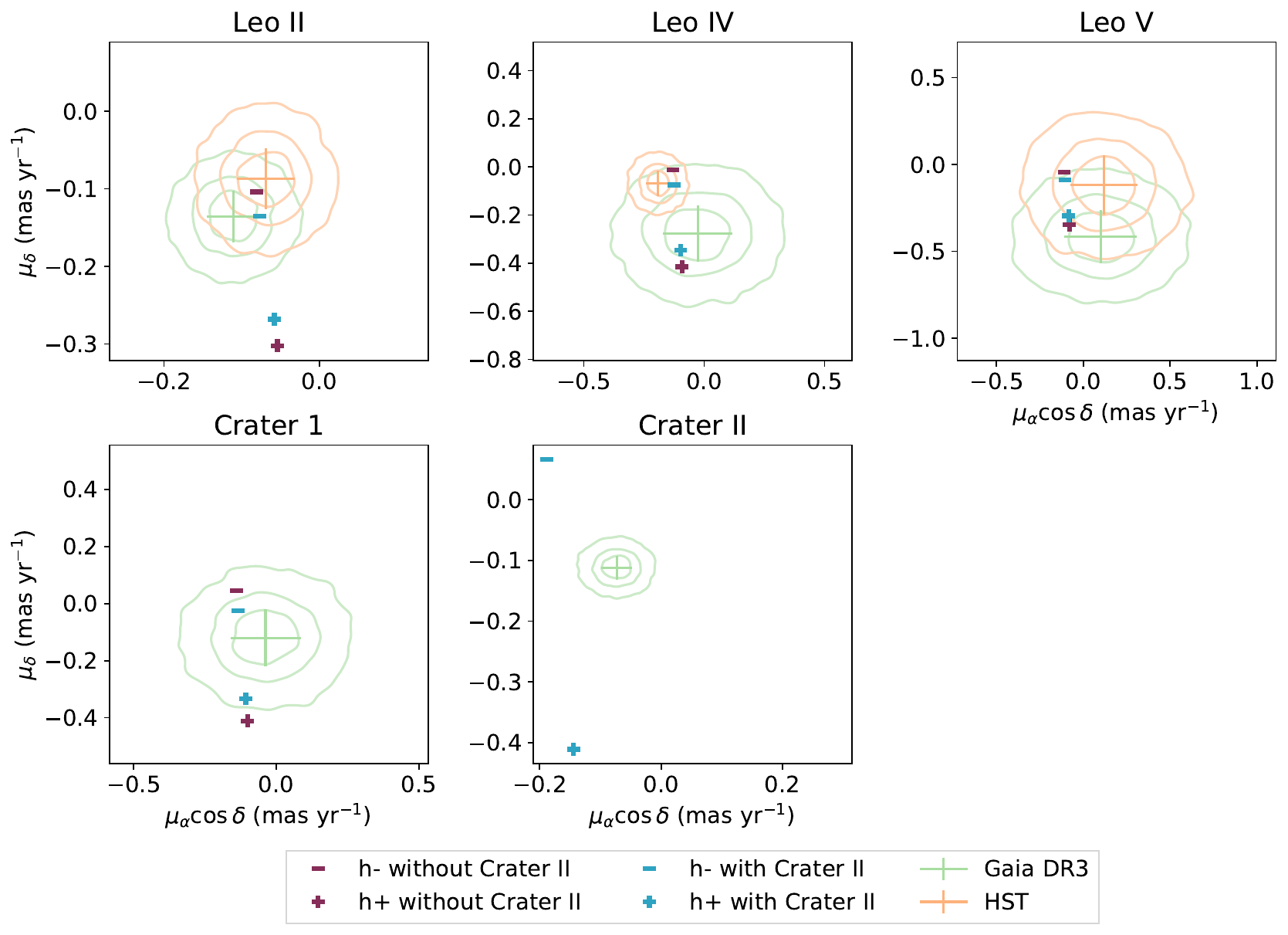}
    \caption{Predicted proper motions of the Crater-Leo objects that we get when we consider the shifts from \cite{garavito-camargo_clustering_2021}, after shifting them back to the current centre of mass of the MW. The magenta "+" and "-" represent the predictions obtained with the negative $h+$ and positive $h-$ circulation about the northern pole, respectively, when we consider Crater II as part of the group. The light-blue "+" and "-" represent the predictions obtained with the negative $h+$ and positive $h-$ circulation about the northern pole, respectively, when we exclude Crater II. The green crosses represent the measured proper motions from \textit{Gaia} DR3 with their uncertainties and the orange crosses represent the proper motion measurements from HST. The corresponding contours correspond to the 34\%, 68\% and 95\% confidence levels for the 2D distribution of the proper motion parameter space.}
    \label{fig:lmc-effect-predictions-total}
\end{figure}

To compare our predictions with the measured proper motions, we shift them back to the reference frame of the MW centre. As before, the prediction for Crater II is far from the measured one, so we can continue to assume that this object was not accreted with the rest of the group, even if the group was indeed accreted before the LMC. The predictions that we get when we include and exclude this dwarf are both represented in Figure~\ref{fig:lmc-effect-predictions-total}. Interestingly, this time, when we consider Crater II as a member, the predicted proper motions that we get in both orbital directions are identical. When we look only at the \textit{Gaia} DR3 measurements, it even looks like Leo II and Leo V might prefer the positive circulation in this scenario. However, their proper motions are so unconstrained that the negative circulation is consistent with $1-2\sigma$ errors. Furthermore, when we remove Crater II, the closest prediction to all the remaining objects is the one that we get when we consider a negative circulation. Since group infall requires a common orbital direction, once again the results lean towards a negative circulation about the northern pole. \par

\section{Discussion and Conclusions}\label{sec:discussion}
We use up-to-date data to infer if the dwarfs Leo II, IV, V, Crater II and the star cluster Crater 1 were accreted as a group into the MW halo, as suggested by \cite{Torrealba2016}. \par  

We present new measurements for the proper motions of Leo IV and Leo V dwarf galaxies using multiepoch HST imaging data. Our final results for the PMs are $(\mu_{\alpha*}, \mu_\delta) = (-0.1921 \pm 0.0514, -0.0686 \pm 0.0523)$ mas yr$^{-1}$ for Leo IV and $(\mu_{\alpha*}, \mu_\delta) = (0.1186 \pm 0.1943, -0.1183 \pm 0.1704)$ mas yr$^{-1}$ for Leo V. \par
We start by determining the orbital poles of the putative group with the proper motions available from \textit{Gaia} DR3, and the new HST proper motions presented in this paper. The possible orbital poles suggest that these objects can share a common value for their specific angular momentum within their uncertainties. When we take into account the new HST proper motion measurements, a common orbital pole becomes even more likely. \par

We use the method proposed by \cite{lynden-bell_ghostly_1995} to predict the proper motions of the considered satellites, assuming that they share a common group origin. Our proper motion predictions are consistent with the measured proper motions within the errors for most of the proposed members. Based on \textit{Gaia} DR3 measurements alone, the predictions of the proper motions for Leo IV and Leo V seem to prefer the positive circulation about the northern pole. In contrast, Leo II and Crater 1 predictions clearly favour the negative circulation. However, a common group accretion requires a common orbiting direction. When we consider our new HST measurements, this changes and all four objects prefer the one co-orbiting direction ($h-$). 

Better proper motion measurements with lower uncertainties are required to confirm this association with higher confidence.

Crater II, however, has its proper motion well constrained and does not share a similar orbital pole with the rest of the objects, nor does it match our proper motion predictions. Furthermore, Crater II is the object with the highest offset from the common value of the line-of-sight shared by the group. We can thus rule out an association with the other members. \par 

Other objects were identified as possible additional members of the Crater-Leo group due to their orbital poles. Yet, their proper motion predictions are far from the measured ones, suggesting that they were not part of it. \par

Based on HST and \textit{Gaia} DR3 proper motions, we have performed orbital analyses for the four satellites that we consider as possibly associated, Leo II, Leo IV, Leo V and Crater 1, to try to understand if they have a shared orbital history or if there is any evidence that the objects have been bound in the past. For both considered potentials, all objects are near their apocentre at the present time. All objects are consistent with having reached their pericentre around the same time, within their uncertainties. When we look at the separation of Leo IV, Leo V and Crater 1 with respect to Leo II, they are all at comparable distances from each other $\sim 3$ Gyr ago, suggesting that there is a physical association between these objects and not just a chance alignment at the present time. The derived orbital properties are consistent, within their observational uncertainties, with those we predict from assuming group infall, especially when we compare with the properties obtained with our new HST proper motions. \par 

We show that the bound orbits derived from the observational uncertainties match the orbits derived from the predicted PMs. These are the orbits that not only have PMs consistent with observations, but that also have similar specific energy and angular momentum. Our assumptions can thus explain why these align along a great circle at the present time. \par

When we consider the effect that the infall of the LMC might have had on the orbital properties of our putative group, it is found that the probability of the objects sharing the same specific angular momentum and specific energy increases. Once again, it seems that Crater II was not accreted with the other satellites. When we shift our predicted proper motions back to the centre of mass of the current MW halo, to compare with the current measurements, we get similar results to the previous ones, suggesting that the infall of the LMC did not have a strong effect on the orbital poles of the considered objects, not surprisingly, as they largely reside on the opposite side of the MW.  The predictions that the proper motions seem to prefer the negative circulation about the northern pole are strengthened when we consider the new proper motion measurements. \par 

Leo II, Leo IV, Leo V, and Crater 1 show orbital properties consistent with those we predict from assuming group infall; however, uncertainties of the current proper motion measurements with both Gaia and HST are too high to conclusively confirm their dynamical associations. The next data releases for Gaia will likely reduce both the systematic and random uncertainties for these objects. Furthermore, with an increased time baseline, proper motion measurements using future HST or JWST observations will provide a definitive answer on whether (some of) the Crater-Leo objects constitute the first identified case of a cosmologically expected, typical group infall event. Time is on our side.

\begin{acknowledgement}
MPJ and MSP acknowledge funding via a Leibniz-Junior Research Group (project number J94/2020). MPJ thanks Gurtina Besla for the helpful advice. MPJ thanks Nuno Humberto for the valuable discussions and suggestions. MPJ thanks the "Summer School for Astrostatistics in Crete" for providing training on the statistical methods adopted in this work. This research has made use of agama \citep{vasiliev_agama_2019}, Astropy \citep{astropy}, corner.py \citep{corner}, galpy \citep{bovy_galpy_2015}, matplotlib \citep{matplotlib}, NASA’s Astrophysics Data System Bibliographic Services, NumPy \citep{numpy} and SciPy \citep{scipy}. Based on observations made with the NASA/ESA Hubble Space Telescope, obtained from the Data Archive at the Space Telescope Science Institute (STScI). Support for this work was provided by NASA through grants for programs GO-14770 and 15507 from STScI, which is operated by the Association of Universities for Research in Astronomy (AURA), Inc., under NASA contract NAS5-26555. These projects are part of the HSTPROMO (High resolution Space Telescope PROper MOtion) Collaboration, a set of projects aimed at improving our dynamical understanding of stars, clusters, and galaxies in the nearby Universe through measurement and interpretation of proper motions from HST, Gaia, and other space observatories. We thank the collaboration members for sharing their ideas and software.
\end{acknowledgement}


\bibliographystyle{aa}
\bibliography{craterleo} 



\appendix

\section{Intrinsic scatter}\label{app:mcmc}
\begin{figure}[ht]
    \centering
    \includegraphics[width=\columnwidth]{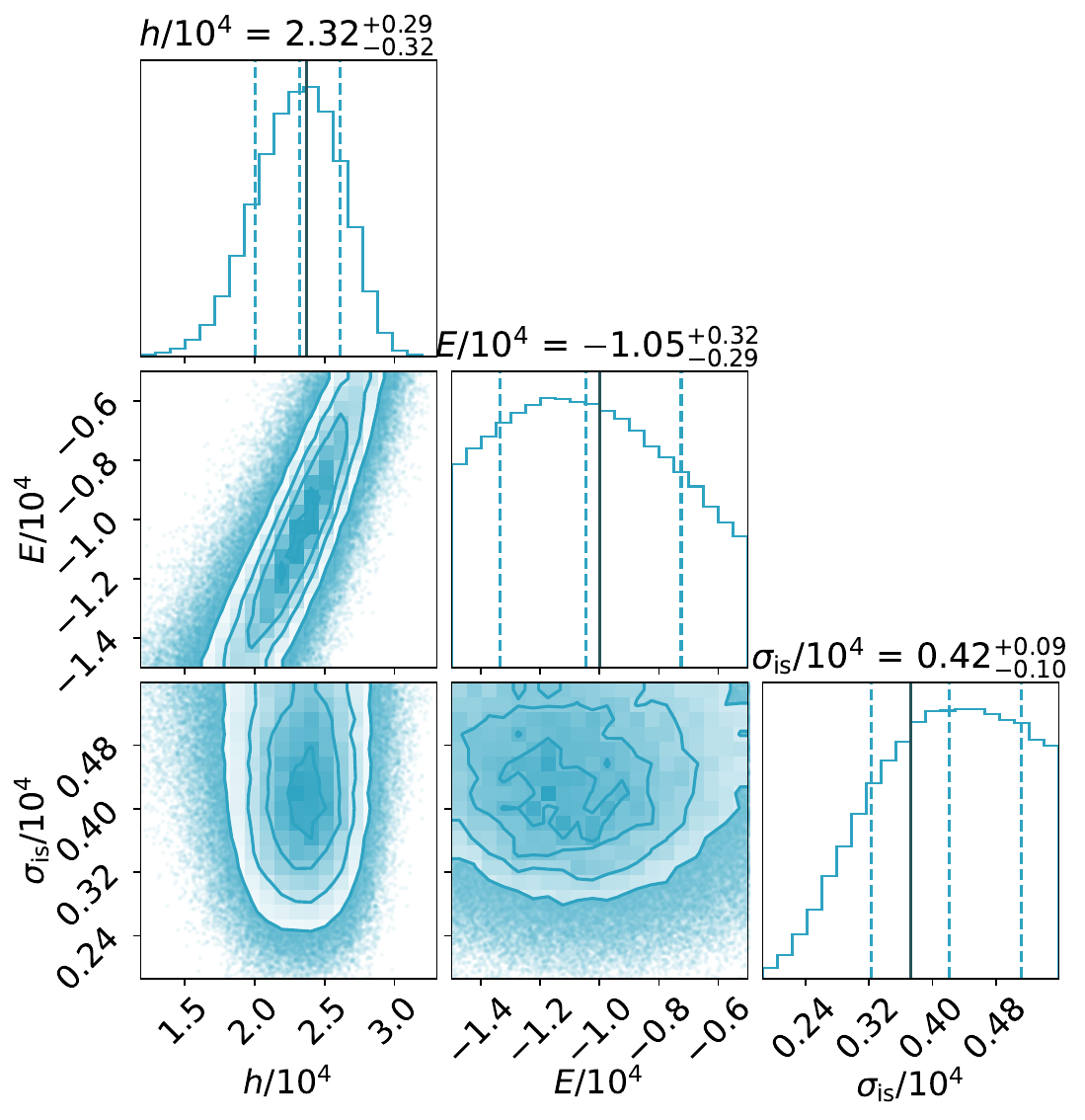}
    \includegraphics[width=\columnwidth]{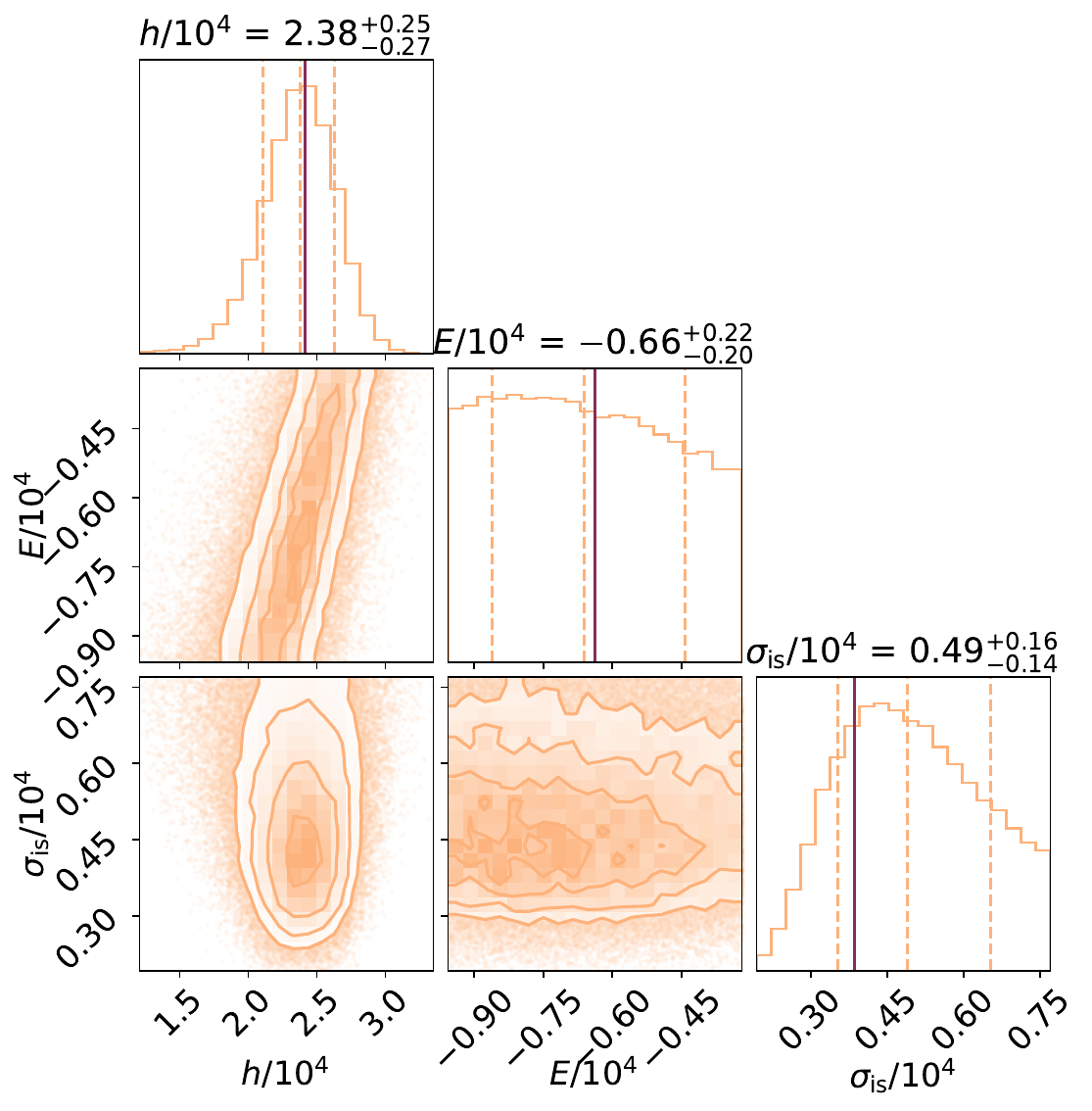}
    \caption{Posterior distributions for $h_0$, $E_0$ and $\sigma_\mathrm{is_0}$ for the original putative group. The histograms along the diagonal represent the posterior distribution for each parameter. Their units are omitted for clarity. The vertical dashed lines indicate the median and 68$\%$ confidence interval. The bottom left panel represents the 2D posterior distribution of these parameters, with the contours corresponding to the $0.5\sigma$, $1\sigma$, $1.5\sigma$, and $2\sigma$ confidence levels, where $\sigma$ is the standard deviation of the 2D distribution. The solid lines represent the values for $h_0$ and $E_0$ that we get from the fit of the Lynden-Bell method. \textbf{Top:} Results considering the MW-only potential. \textbf{Bottom:} Results considering the MW+LMC potential.}
    \label{fig:cornerplots-group}
\end{figure}

\begin{figure}[ht]
    \centering
    \includegraphics[width=\columnwidth]{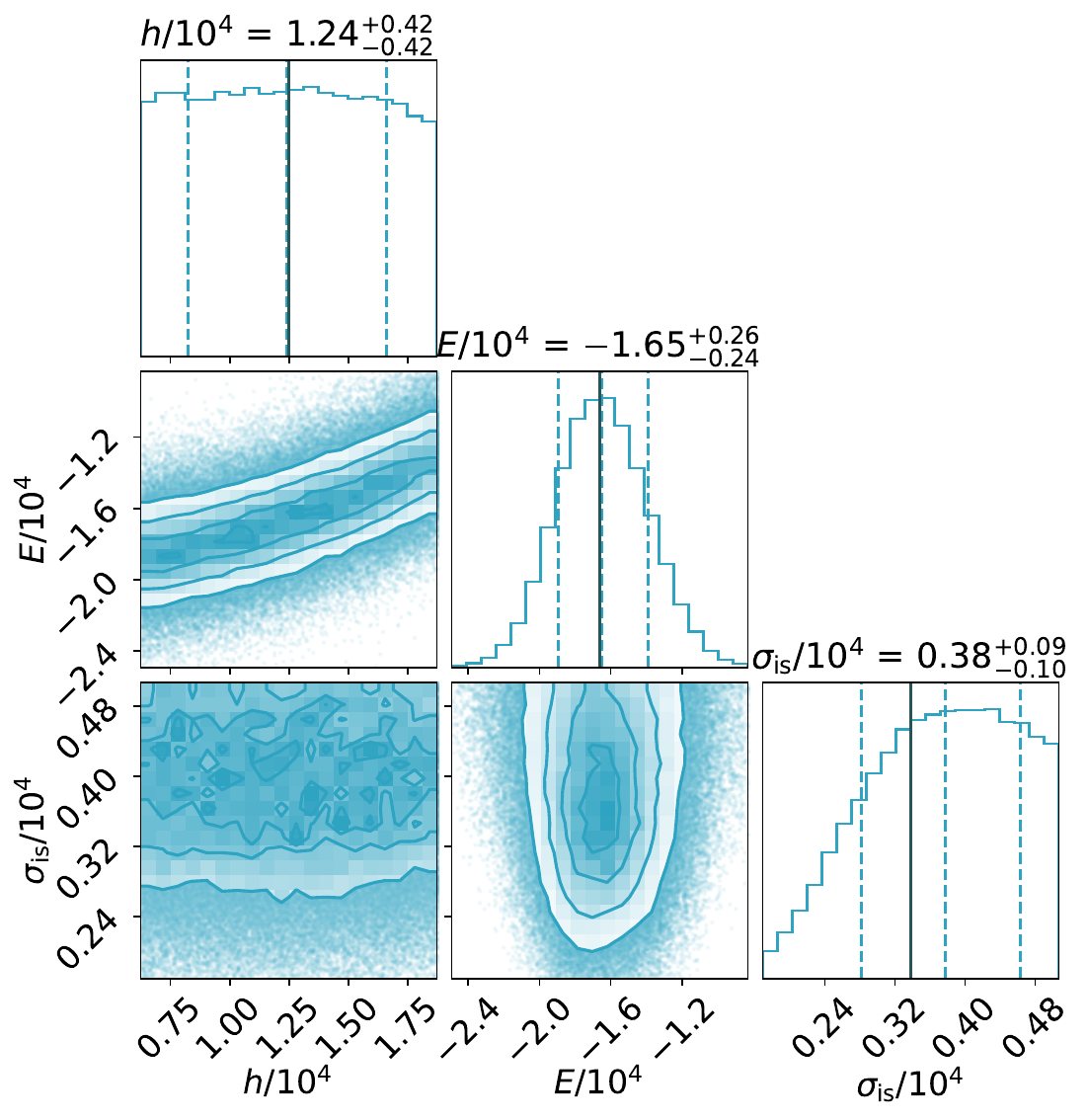}
    \includegraphics[width=\columnwidth]{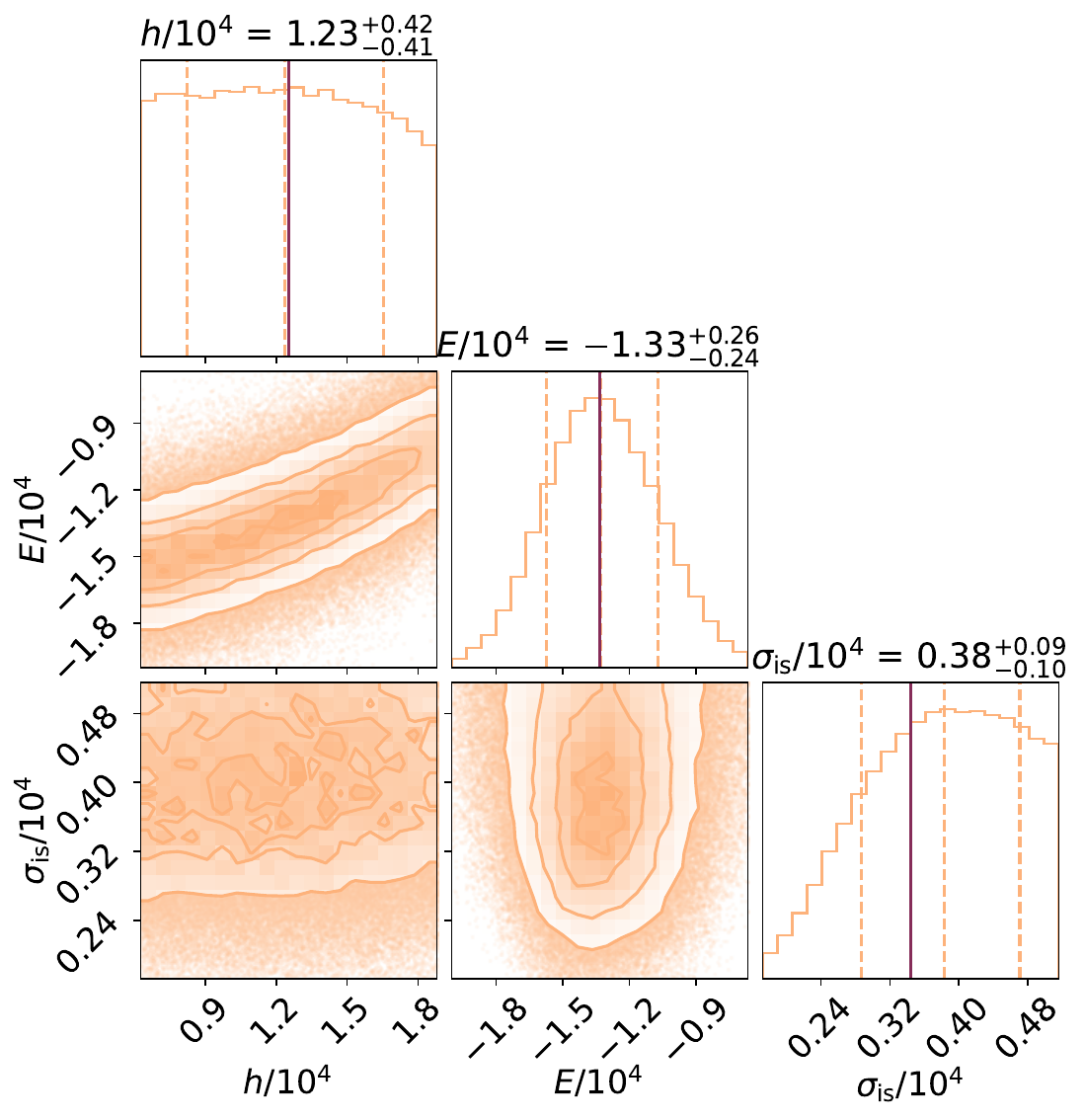}
    \caption{Posterior distributions for $h_0$, $E_0$ and $\sigma_\mathrm{is_0}$ for the putative group after removing Crater II. The histograms along the diagonal represent the posterior distribution for each parameter. Their units are omitted for clarity. The vertical dashed lines indicate the median and 68$\%$ confidence interval. The bottom left panel represents the 2D posterior distribution of these parameters, with the contours corresponding to the $0.5\sigma$, $1\sigma$, $1.5\sigma$, and $2\sigma$ confidence levels, where $\sigma$ is the standard deviation of the 2D distribution. The solid lines represent the values for $h_0$ and $E_0$ that we get from the fit of the Lynden-Bell method. \textbf{Top:} Results considering the MW-only potential. \textbf{Bottom:} Results considering the MW+LMC potential.}
    \label{fig:cornerplots-group-without-craterii}
\end{figure}

The intrinsic scatter of the putative group was determined by the method described in Sect.~\ref{sec:mcmc}. Here, the posterior distributions of $h_0$, $E_0$ and $\sigma_\mathrm{is_0}$ are represented for both potentials (MW-only, in blue and MW+LMC, in orange), for both the cases with (Fig.~\ref{fig:cornerplots-group}) and without Crater II (Fig.~\ref{fig:cornerplots-group-without-craterii}).

\section{Effect of the MW mass}\label{app:mwmass}
\begin{figure}[ht]
    \centering
    \includegraphics[width=\columnwidth]{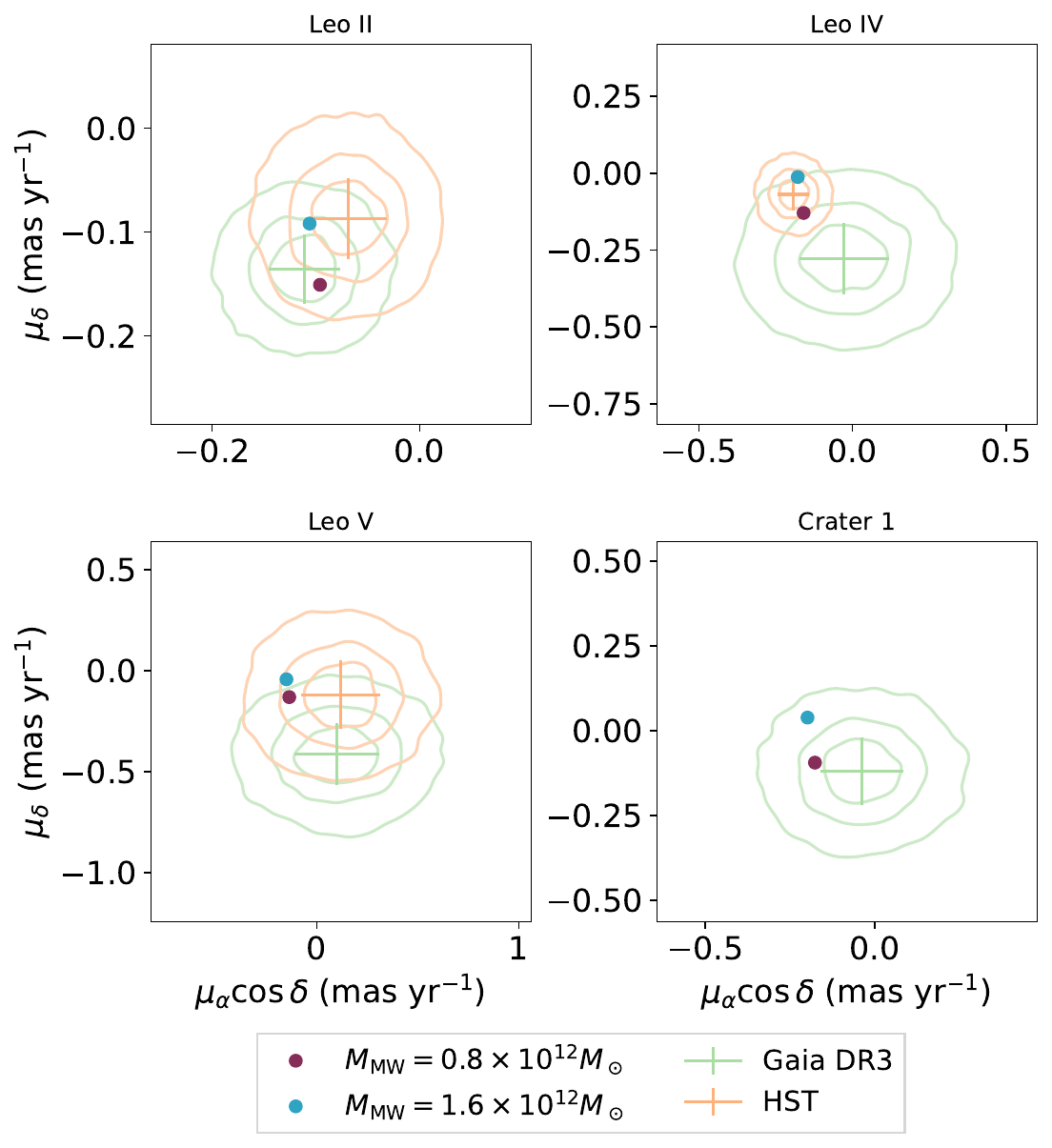}
    \caption{Predicted proper motions of the Crater-Leo objects for different masses of the MW. The magenta and blue dots represent the predictions obtained with $M_\mathrm{MW} = 0.8\times10^{12}M_\odot$ and $M_\mathrm{MW} = 1.6\times10^{12}M_\odot$, respectively. For both we assume the negative $h-$ circulation about the northern pole. The green crosses represent the measured proper motions from \textit{Gaia} DR3 with their uncertainties and the orange crosses represent the measured proper motions from HST. The contours correspond to the 34\%, 68\% and 95\% confidence levels for the 2D distribution of the proper motion parameter space.}
    \label{fig:predicted-pm-different-potentials}
\end{figure}

Since the mass of the MW is still unconstrained, here, we check what happens to the predicted proper motions when we increase its mass. As can be seen in Figure~\ref{fig:predicted-pm-different-potentials}, the predictions slightly change but the conclusions remain the same. We decided to plot only the proper motions obtained with the negative $h-$ direction since our whole analyses show that this is the preferred direction of the group. 

\section{Additional members}\label{app:additional-members}
The possible additional members of the group were determined by the previously described method, in Sect.~\ref{sec:pudim-pt2}, taking into account their orbital pole positions. Here, we can find the comparison between the predictions that we get for each object, assuming a group with Crater 1, Leo II, Leo IV and Leo V, plus one of the possible additional members. For simplicity, only the predictions for the additional members are displayed.

\begin{figure}[ht]
    \centering
    \includegraphics[width=\columnwidth]{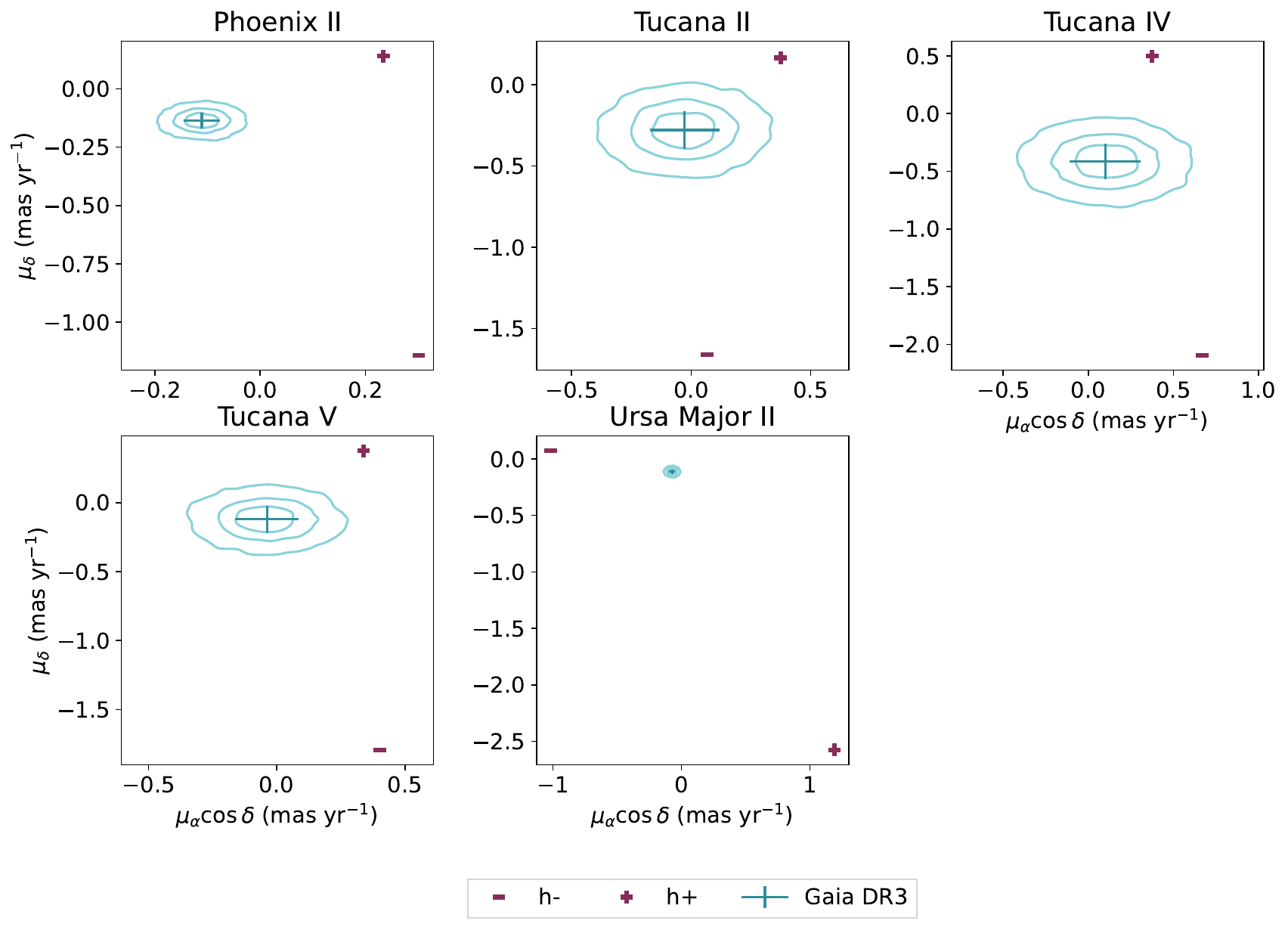}
    \caption{Predicted proper motions of the possible additional members of the Crater-Leo group. The light-blue "+" and dark-blue "-" represent the predictions obtained with the negative $h+$ and positive $h-$ circulation about the northern pole, respectively. The green crosses represent the measured proper motions from \textit{Gaia} DR3 with their uncertainties. The contours correspond to the 34\%, 68\% and 95\% confidence levels for the 2D distribution of the proper motion parameter space.}
    \label{fig:predicted-pm-additional-members}
\end{figure}


\section{Orbital histories}\label{app:orbital}
In Sect.~\ref{sec:orbital-parameters} we discussed the orbital histories of the Crater-Leo objects for both the MW-only and the MW-LMC potential. Here, we show supplementary figures to support the results discussed in the main text. 

\begin{figure*}
    \centering
    \includegraphics[scale=0.4]{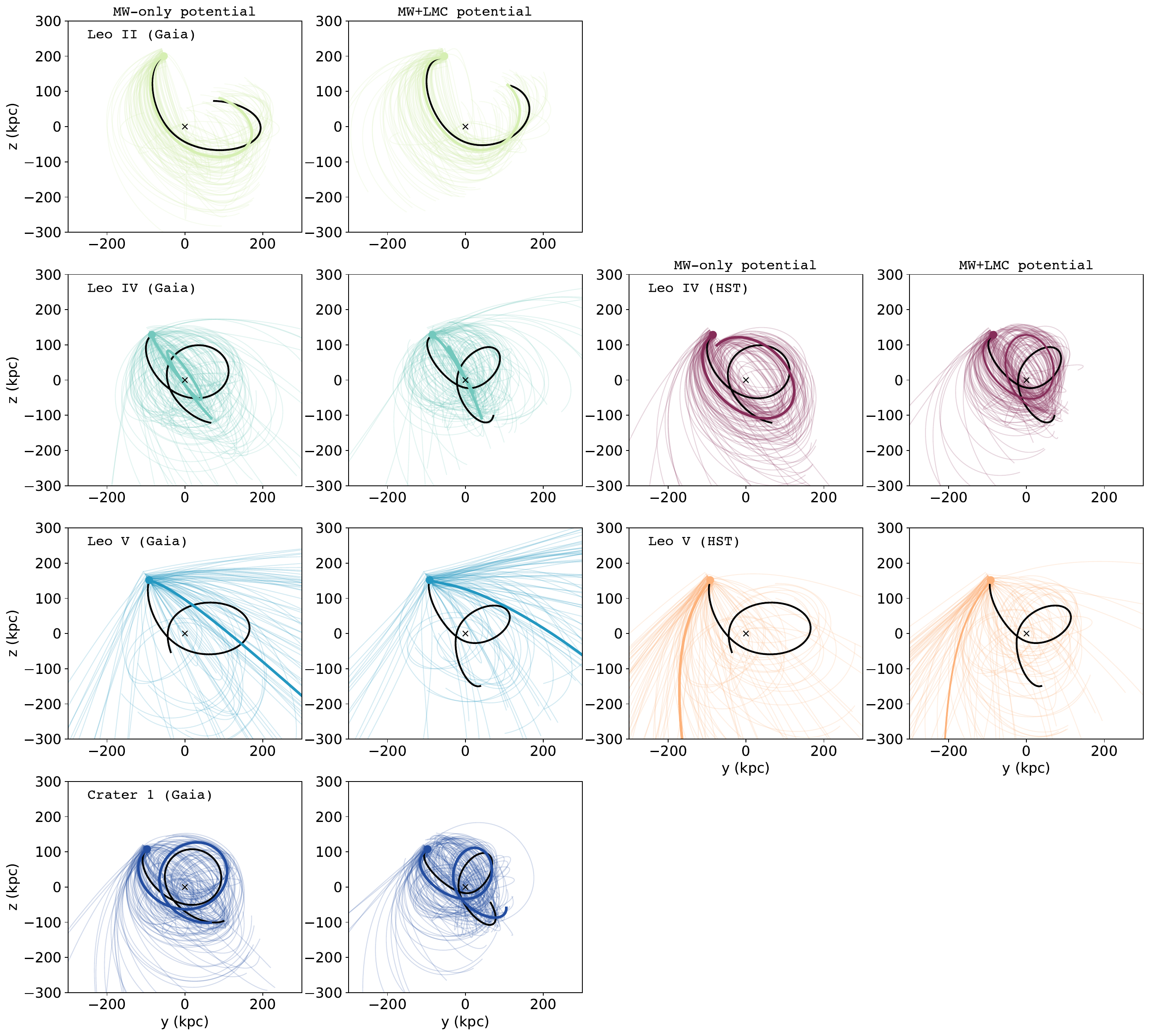}
     \caption{Orbital evolution for the proposed group up to 6 Gyr in the Galactocentric Cartesian $Y-Z$ plane using both the MW-only potential and the MW+LMC potential. The thick lines indicate their most likely orbit and the thinner lines represent 100 Monte-Carlo realisations. The black solid line represents the orbital evolution obtained using the predicted proper motions for each object.}
    \label{fig:face-on-mc}
\end{figure*}

\section{Effect of the LMC}\label{app:lmceffect}

\begin{table}[ht]
	\centering
	\caption{New positions in Galactocentric coordinates and velocities of the Crater-Leo objects after applying the shifts from \cite{garavito-camargo_clustering_2021}.}
	\label{tab:lmc_effect_shifts}
	\begin{tabular}{lcccccc} 
		\hline
		  Object & $x$ & $y$ & $z$ & $v_x$ & $v_y$ & $v_z$ \\
                   & \multicolumn{3}{c}{(kpc)} & \multicolumn{3}{c}{(kms$^{-1}$)} \\
		\hline
		Leo II & $-68.4$ & $-35.3$ & $185.2$ & $-48.3$ & $67.4$ & $-22.5$    \\
		Leo IV & $-13.2$ & $-78.7$ & $116.6$ & $92.6$ & $16.5$ & $-24.0$   \\
		Leo V  & $-18.8$ & $-81.7$ & $138.2$ & $246.0$ & $-76.8$ & $9.9$  \\
            Crater 1 & $1.5$ & $-92.3$ & $96.3$ & $42.6$ & $83.4$ & $20.8$  \\
            Crater II & $12.2$ & $-82.7$ & $69.1$ & $23.6$ & $127.8$ & $-22.5$ \\
		\hline
	\end{tabular} 
\end{table}

\begin{figure}[ht]
    \centering
    \includegraphics[width=\columnwidth]{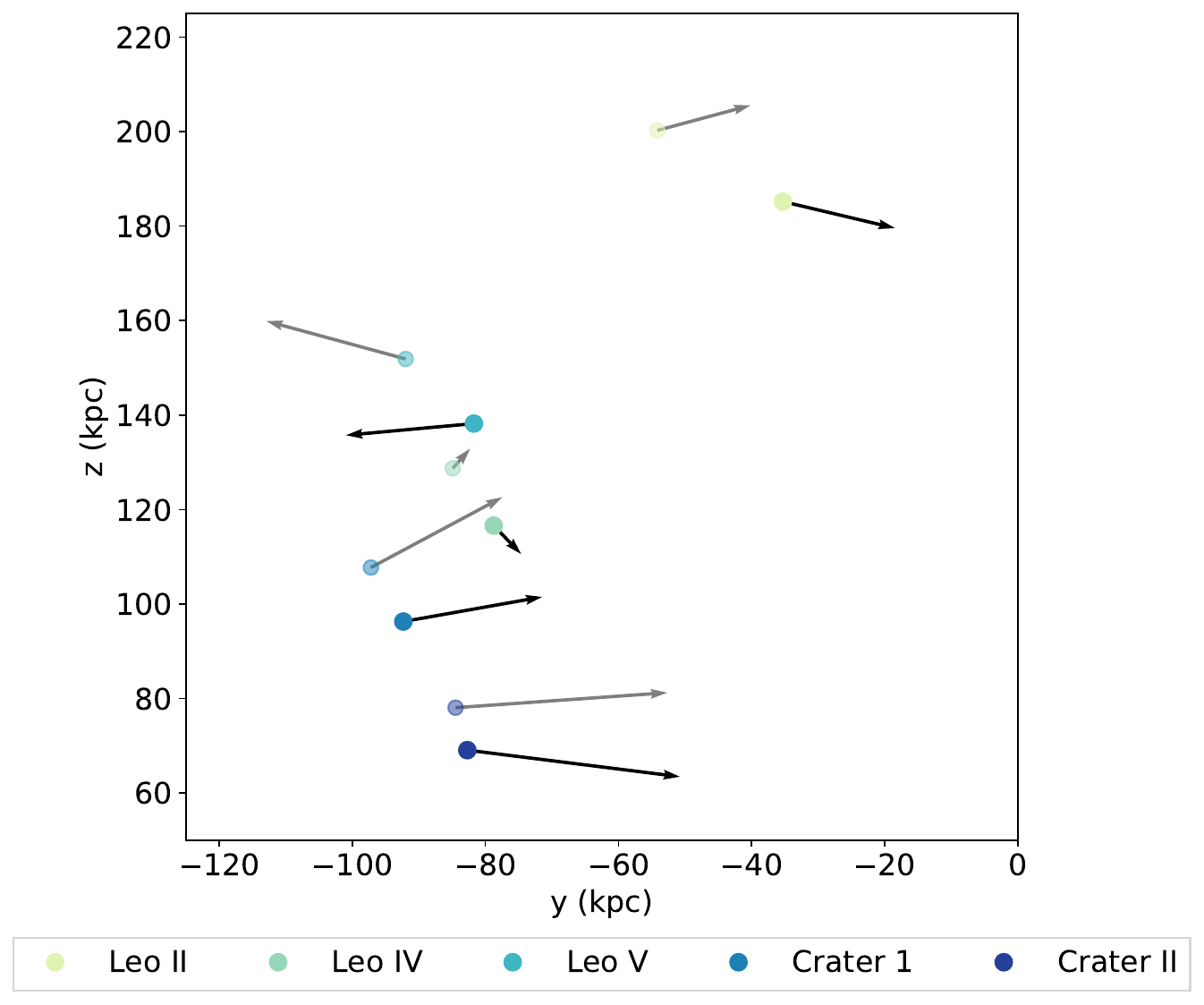}
    \caption{Positions and velocities (as vectors) of the Crater-Leo objects. The bigger markers correspond to the position/velocity after applying the shifts from \cite{garavito-camargo_clustering_2021} and the smaller and fainter markers indicate the current measurements.}
    \label{fig:lmc-effect-shifts}
\end{figure}

\begin{figure}[ht]
    \centering
    \includegraphics[width=\columnwidth]{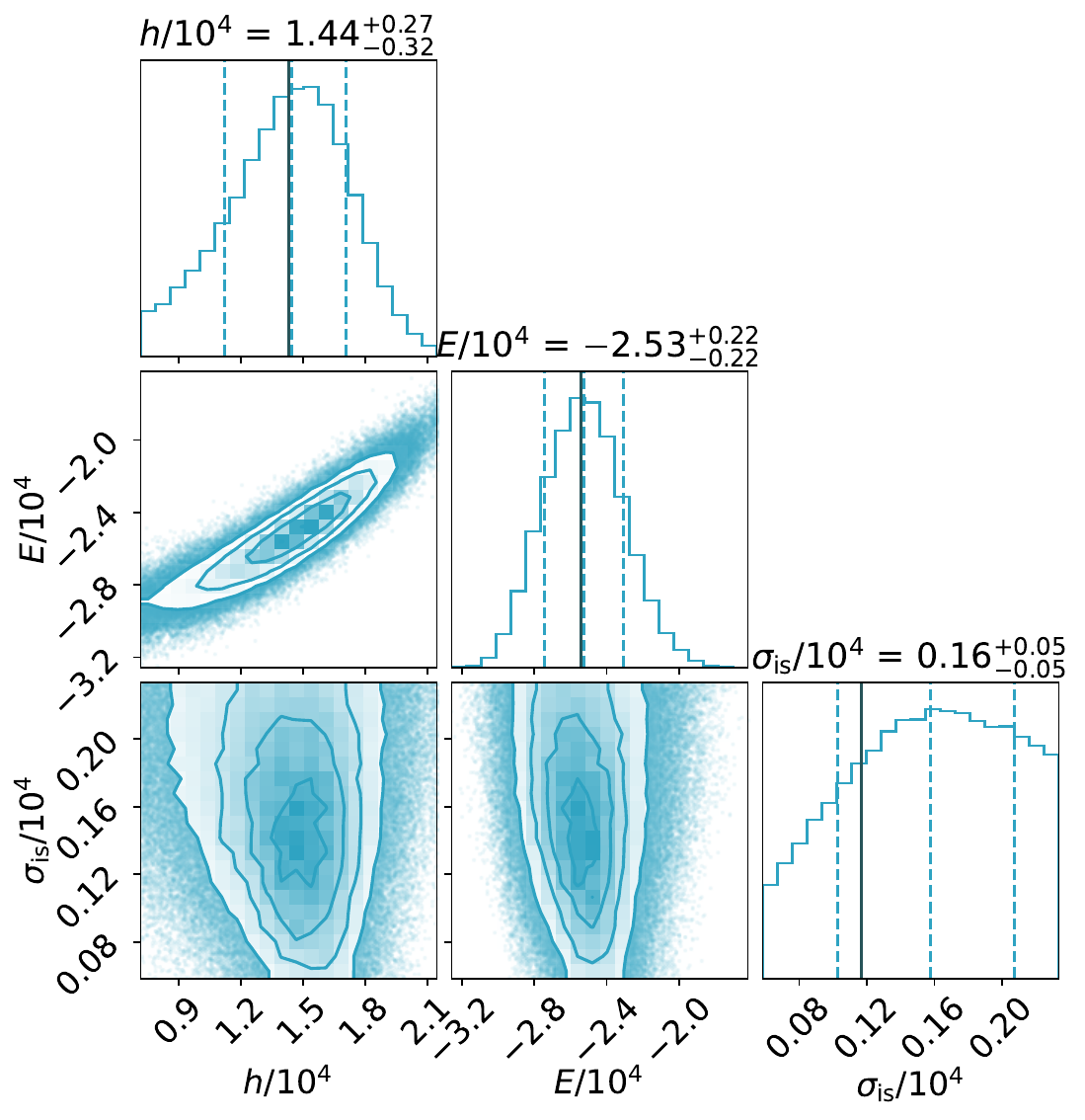}
    \includegraphics[width=\columnwidth]{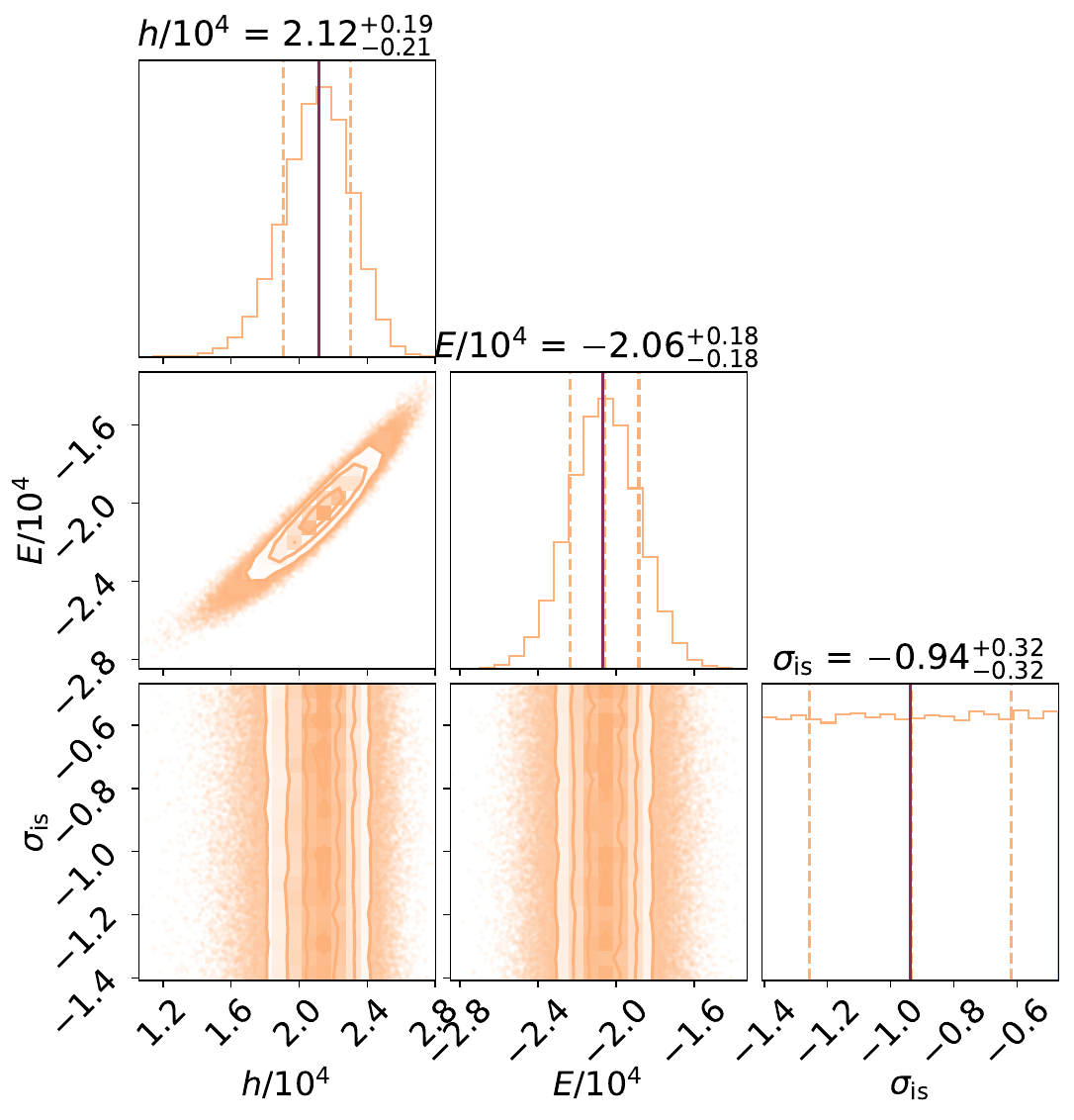}
    \caption{Posterior distributions for $h_0$, $E_0$ and $\sigma_\mathrm{is_0}$ for the putative group considering the LMC effect. The histograms along the diagonal represent the posterior distribution for each parameter. Their units are omitted for clarity. The vertical dashed lines indicate the median and 68$\%$ confidence interval. The bottom left panel represents the 2D posterior distribution of these parameters, with the contours corresponding to the $0.5\sigma$, $1\sigma$, $1.5\sigma$, and $2\sigma$ confidence levels, where $\sigma$ is the standard deviation of the 2D distribution. The solid lines represent the values for $h_0$ and $E_0$ that we get from the fit of the Lynden-Bell method. \textbf{Top:} Results considering Crater II as part of the group. \textbf{Bottom:} Results excluding Crater II from the group.}
    \label{fig:cornerplots-lmc-effect}
\end{figure}

\end{document}